\documentclass{article}
\usepackage[utf8]{inputenc}
\usepackage{graphicx}
\usepackage[figuresleft]{rotating}
\usepackage{amsmath}
\usepackage{bm}
\usepackage[margin=1in]{geometry}
\usepackage{threeparttable}
\usepackage{booktabs}
\usepackage{tabularx}
\usepackage{url}
\usepackage{hyperref}
\usepackage{caption}
\usepackage{natbib}
\usepackage[english]{babel}
\usepackage[affil-it]{authblk}
\usepackage{comment}
\usepackage{titlesec}
\usepackage{textcomp}

\titleformat*{\subsection}{\itshape}

\title{Probabilistic Projections of Baseline 21st Century CO$_2$ Emissions Using a Simple Calibrated Integrated Assessment Model}
\date{}
\author[1]{Vivek Srikrishnan}
\author[2]{Yawen Guan}
\author[3,4,5]{Richard S. J. Tol}
\author[6,7,8]{Klaus Keller}
\affil[1]{Department of Biological \& Environmental Engineering, Cornell University, Ithaca, NY, USA}
\affil[2]{Department of Statistics, University of Nebraska, Lincoln, NE, USA}
\affil[3]{Department of Economics, University of Sussex, Sussex, UK}
\affil[4]{Institute for Environmental Studies, Vrije Universiteit Amsterdam, Amsterdam, The Netherlands}
\affil[5]{Department of Spatial Economics, Vrije Universiteit Amsterdam, Amsterdam, The Netherlands}
\affil[6]{Department of Geosciences, Pennsylvania State University, University Park, PA, USA}
\affil[7]{Earth and Environmental Systems Institute, Pennsylvania State University, University Park, PA, USA}
\affil[8]{Thayer School of Engineering, Dartmouth College, Hanover, NH, USA}

\begin{document}

\maketitle

%\begin{translation}{english}
\begin{abstract}
Probabilistic projections of baseline (with no additional mitigation policies) future carbon emissions are important for sound climate risk assessments. Deep uncertainty surrounds many drivers of projected emissions. Here we use a simple integrated assessment model, calibrated to century-scale data and expert assessments of baseline emissions, global economic growth, and population growth, to make probabilistic projections of carbon emissions through 2100. Under a variety of assumptions about fossil fuel resource levels and decarbonization rates, our projections largely agree with several emissions projections under current policy conditions. Our global sensitivity analysis identifies several key economic drivers of uncertainty in future emissions and shows important higher-level interactions between economic and technological parameters, while population uncertainties are less important. Our analysis also projects relatively low global economic growth rates over the remainder of the century. This illustrates the importance of additional research into economic growth dynamics for climate risk assessment, especially if pledged and future climate mitigation policies are weakened or have delayed implementations. These results showcase the power of using a simple, transparent, and calibrated model. While the simple model structure has several advantages, it also creates caveats for our results which are related to important areas for further research.
\end{abstract}
%\end{translation}

%\linenumbers
\section{Introduction}

What is a sound approach to projecting future climate change and its impacts? This is a critical question for understanding the impact of adaptation and mitigation strategies. Projected climatic changes hinge on Earth system properties and future drivers, including anthropogenic carbon dioxide (CO$_2$) emissions. Projections of future anthropogenic CO$_2$ emissions are deeply uncertain; there is no consensus about the probability distribution of future emissions~\citep{hoNotAllCarbon2019}. Poorly calibrated emissions projections translate into poor climate projections and contribute to poor risk management decisions~\citep{morganImprovingWayWe2008}.

One approach, adopted in the reports of the Intergovernmental Panel on Climate Change, or IPCC~\citep{ipccClimateChange20142014}, to handling the deep uncertainty associated with future emissions and the associated radiative forcing is to use scenarios which cover an appropriate range of plausible futures. For example, the Shared Socioeconomic Pathways, or SSPs~\citep{oneillNewScenarioFramework2014a, riahiSharedSocioeconomicPathways2017} explore a plausible range of future emissions through an internally consistent set of future socioeconomic narratives. These scenarios are useful for creating a set of harmonized assumptions for modeling and impacts studies. As they are not intended to be interpreted as predictions of future socioeconomic, emissions, or climate trajectories, they are explicitly provided without probabilities or likelihoods~\citep{vanvuurenRepresentativeConcentrationPathways2011}. This framework is useful for many purposes. However, this lack of probabilistic information makes them difficult to integrate into risk assessment for climate change impacts or adaptation and makes their interpretation susceptible to the cognitive biases that interfere with decision-making under uncertainty~\citep{tverskyJudgmentUncertaintyHeuristics1974,morganUncertaintyGuideDealing1992, websterUncertaintyEmissionsProjections2001}. For example, one problematic approach is to view all scenarios as equally likely~\citep{wigleyInterpretationHighProjections2001a}.

Another more recent example is the controversy over presentations of the highest Representative Concentration Pathway, RCP 8.5~\citep{riahiRCPScenarioComparatively2011}, as a baseline scenario for impacts~\citep{hausfatherEmissionsBusinessUsual2020}. We refer a ``baseline scenario'' as one with no inclusion of the effects of mitigation policies beyond those currently implemented, rather than the original use of the term in the integrated assessment modeling (IAM) literature, where ``baseline'' refers specifically to a scenario generated by a model run with no forced mitigation through climate policies~\citep{vanvuurenRepresentativeConcentrationPathways2011}. Critics of the interpretation that RCP 8.5 represents a baseline (no inclusion of additional mitigation policies beyond those currently implemented) radiative forcing pathway raise concerns about the increase in coal energy share, relative to present trends, required to achieve this forcing by IAMs~\citep{ritchie1000GtCCoal2017, ritchieWhyClimateChange2017, hausfatherEmissionsBusinessUsual2020, burgessIPCCBaselineScenarios2021}. Indeed, as noted by \citet{hausfatherEmissionsBusinessUsual2020} and \citet{skeaOutlooksExplorationsNormative2021}, current International Energy Agency (IEA) projections diverge from the emissions required by higher-emitting scenarios such as RCP 8.5, and \citet{oneillAchievementsNeedsClimate2020} list keeping scenarios up to date and adding additional scenarios in more risk-relevant areas of the scenario space as two key future needs for scenario development. On the other hand, \citet{schwalmRCP8TracksCumulative2020} observes that the emissions trajectory associated with the older standalone RCP 8.5 (rather than the newer joint SSP5-8.5 scenario~\citep{krieglerNewScenarioFramework2014}) closely tracks recent emissions, though this arises from differences between projected and observed land-use change emissions rather than fossil emissions~\citep{hausfatherRCP8ProblematicScenario2020}. Additionally, along with RCP 4.5, RCP 8.5 bounds a reasonable range surrounding projected emissions through 2050, particularly when land-use change emissions are included. As a result, the authors conclude that RCP 8.5 continues to provide value as a short-to-medium term mean scenario and a longer-term tail-risk scenario.

Both of these seemingly antagonistic perspectives are driven and justified by different use cases and information needs. In this case, the epistemic value provided by RCP 8.5 depends on the question posed and if the goal is to identify a range of illustrative outcomes or to enumerate probabilities of outcomes to assess and manage risk. This latter consideration is complicated by the underlying deep uncertainties. For example, there is little consensus on key drivers such as projections of future economic growth~\citep{christensenUncertaintyForecastsLongrun2018} or the penetration rates of zero-carbon technologies in the global energy mix. Uncertainty about the future strength and direction of mitigation policies further complicates the calculation of probabilities. The potential role of negative emissions technologies (NETs) is another deep uncertainty that directly affects projected emissions trajectories. These considerations demonstrate the need for a transparent and systematic understanding of the relevant assumptions and dynamics and underpin a particular probabilistic projection, so analysts and decision-makers can better understand and account for caveats associated with the resulting probabilities. 

A key consideration in producing probabilistic projections is the trade-off between two competing modeling objectives: realistic dynamics to capture key processes and utilize interpretable parameters versus sufficiently fast model evaluations to enable careful uncertainty characterization and quantification. The previous literature on probabilistic emissions projections demonstrates a variety of approaches to navigating this trade-off, which are motivated by the underlying research question. When highly-detailed, computationally-expensive ``bottom-up'' models are used, such as the Global Change Assessment Model (GCAM) or the Emissions Prediction and Policy Analysis model (EPPA), one option is to focus on the influence of a few uncertain inputs, as this uncertainty space can be captured with a relatively small-to-moderate number of samples \citep[\emph{e.g.}][]{capellan-perezLikelihoodClimateChange2016, fykeProbabilisticAnalysisCumulative2015}. One downside to this approach with complex models is that the impact of interactions between the inputs treated as uncertain and those which are not may be missed. \citet{vanvuurenConditionalProbabilisticEstimates2008} use a broader set of inputs, but focus on sampling around each of the no-policy IPCC-SRES scenarios rather than sampling the entire probability space. A larger set of samples can be used with an emulator of the full model~\citep[\emph{e.g.}][]{websterUncertaintyEmissionsProjections2002, sokolovProbabilisticForecastTwentyFirstCentury2009, websterAnalysisClimatePolicy2012, gillinghamModelingUncertaintyIntegrated2018a}, as emulation allows for many more model evaluations within a fixed computational budget, resulting in an ability to explore many more sources of uncertainty at the expense of losing some of the dynamical richness of the full model (depending on the type of emulator or response surface used).  \citet{gillinghamModelingUncertaintyIntegrated2018a} go further in building emulators of multiple IAMs, which are used to understand the impact of model structural uncertainty on the resulting projections. Another approach is the use of a Bayesian statistical model calibrated using historical data~\citep[\emph{e.g.}][]{rafteryLessWarming21002017, liuCountrybasedRateEmissions2021}. These models can also be run many times, allowing them to fully resolve the tails of the projective distributions, and are flexible enough to capture historical dynamics while representing different future scenarios and potential trend breaks. However, they may have parameters which are less interpretable, potentially resulting in fewer insights from examining sensitivities and marginal parameter distributions. 

This spread of approaches is valuable as it reveals the impacts of the underlying modeling assumptions and included uncertainties on the resulting projections. In this study, we add to this literature by using a simple, mechanistically-motivated integrated assessment model. The simplicity of the model permits millions of model evaluations without requiring a reduced-form emulator, allowing full statistical calibration of the model on historical data and insights into the dynamics of the full model. This level of simplicity provides epistemic benefits by making full uncertainty quantification computationally tractable~\citep{helgesonWhySimplerComputer2021}, much like the statistical approach of \citet{rafteryLessWarming21002017} and \citet{liuCountrybasedRateEmissions2021}. The mechanistically-motivated model structure allows us to incorporate theoretical insights about the dynamics and structure of the relationships between interpretable model parameters using prior distributions drawn from the literature.

We focus on projecting baseline CO$_2$ emissions for the remainder of the 21st century, only incorporating the effects of those mitigation policies which have had sufficient effect to be reflected in the calibration data (both historical observations and expert assessments made under the baseline assumption). These baseline emissions projections will not necessarily be consistent with the 2010 baselines adopted by the SSPs, as we calibrate, constrain, and initialize our model using more recent observations which have been influenced by past and current policies, as well as expectations about potential future policies. We also neglect the impact of currently unproven but potentially impactful NETs, as it is unclear what technologies might eventually penetrate on a wide scale, when they might do so, and what their level of negative emissions may be \citep{fussBettingNegativeEmissions2014, smithBiophysicalEconomicLimits2015, vaughanExpertAssessmentConcludes2016}. It is also unclear as to whether NETs would be able to achieve a critical deployment level in the absence of additional climate policies~\citep{honeggerPoliticalEconomyNegative2018}, which we do not consider. As a result, these projections are best understood as a reflection of what CO$_2$ emissions might look like without significant changes in mitigation policy implementation or technological development and deployment patterns. We refrain from projection global mean temperatures, as we do not consider the effects of carbon-cycle and biogeochemical dynamics and their uncertainties, which can have a large impact on the resulting CO$_2$ concentrations~\citep{boothNarrowingRangeFuture2017, quilcailleUncertaintyProjectedClimate2018a}, as well as uncertainties related to climate sensitivity~\citep{goodwinBayesianEstimationEarth2021}.

Naturally, the simple structure of our model does mean that some assumptions about the structure are particularly influential, which creates a number of caveats. This analysis illustrates some of the challenges faced in navigating the simplicity-realism trade-off, while also showing the potential for what can be learned using this approach.

\section{Modeling Overview}

In this section, we provide a brief overview of the structure of our model. Full details are available in Section S1 of Online Resource 1. As a starting point, we adopt the overall structure of the DICE model~\citep{nordhausDICE2013RIntroduction2013a, nordhausRevisitingSocialCost2017}. While our model structure is similar to DICE, our analysis differs from one of the typical uses of that model~\citep[\emph{e.g.}][]{nordhausOptimalTransitionPath1992} as we do not optimize over the space of abatement policies. We expand on the DICE model structure by allowing population growth to be endogenous and affected by economic growth. We also use a different approach to represent changes in the emissions intensity of the global economy, which in our model is the result of successive penetrations of technologies with varying emissions intensities. The resulting model structure involves a logistic population growth component with an uncertain saturation level. We model global economic output using a Cobb-Douglas production function in a Solow-Swan model of economic growth. Population and economic output influence each other through changes in per-capita consumption and labor inputs.

Economic output is translated into emissions using a mixture of four emitting technologies: a zero-carbon pre-industrial technology, a high-carbon intensity fossil fuel technology (representative of coal), a lower-carbon intensity fossil fuel technology (representative of oil and gas), and a zero-carbon advanced technology (representative of renewables and nuclear). The lack of NETs in our modeling framework means that we cannot fully explore the bottom of the emissions range captured by the SSP-RCP framework, as several of these scenarios, namely SSP1-1.9, SSP2-2.6, and SSP4-3.4, include their penetration prior to 2100.

We treat all model parameters, including emissions technology penetration dynamics, as uncertain. The statistical model accounts for cross-correlations across the model errors of the three modules as well as independent observation errors. We calibrate the model using century-scale observations of population and global domestic products per capita~\citep{boltMaddisonStyleEstimates2020} and anthropogenic CO$_2$ emissions (excluding land use emissions)~\citep{bodenGlobalRegionalNational2017, friedlingsteinGlobalCarbonBudget2020}. As the \citet{boltMaddisonStyleEstimates2020} data extend only to 2018, we extend them to 2020 using World Bank data~\citep{theworldbankGDPPPPConstant2020} for global domestic product per capita and United Nations data for population~\citep{unitednationsdepartmentofeconomicandsocialaffairspopulationdivisionProbabilisticPopulationProjections2019}. We also probabilistically invert~\citep{kraanUncertaintyCompartmentalModels2000, fullerProbabilisticInversionExpert2017} three expert assessments in our calibration procedure to gain additional information about prior distributions and potential future changes to population~\citep{unitednationsdepartmentofeconomicandsocialaffairspopulationdivisionProbabilisticPopulationProjections2019}, economic output~\citep{christensenUncertaintyForecastsLongrun2018}, and CO$_2$ emissions~\citep{hoNotAllCarbon2019} which are not reflected in the historical data. More information about the model calibration procedure is available in Section S2 of Online Resource 1, while full details on the derivation of the likelihood function and the choice of prior distributions are available in Sections S3 and S4 of Online Resource 1, respectively.

Two key deep uncertainties affecting probabilistic projections of CO$_2$ emissions are (i) the size of the fossil fuel resource base~\citep{capellan-perezLikelihoodClimateChange2016, wangImplicationsFossilFuel2017} and (ii) the prior beliefs about the penetration rate of zero- or low-carbon energy technologies~\citep{gambhirAssessingFeasibilityGlobal2017}. First, we consider the impact of unknown quantities of remaining fossil fuel resources, which are fossil fuel deposits which are potentially recoverable (as opposed to reserves, which are available for profitable extraction with current prices and technologies)~\citep{rognerAssessmentWorldHydrocarbon1997}. Estimates of the fossil fuel resource base vary widely~\citep{mcgladeReviewUncertaintiesEstimates2012, mcgladeMethodsEstimatingShale2013, mohrProjectionWorldFossil2015, ritchie1000GtCCoal2017}. Recent criticisms of the continued use of high-emissions and high-forcings scenarios focus on the plausibility of the required amount of fossil fuels, particularly coal, required to generate the emissions associated with these scenarios~\citep{ritchie1000GtCCoal2017, ritchieWhyClimateChange2017}. For the penetration rate of low-carbon technologies, historical emissions data can only provide limited information about this transition due to the relatively limited penetration of these technologies to date and differences in the penetration dynamics of other generating technologies such as coal, oil, and natural gas~\citep{gambhirAssessingFeasibilityGlobal2017}. 

We focus on these deep uncertainties to illustrate the sensitivity of probabilistic projections of emissions, and therefore temperature anomalies, to these assumptions, while recognizing that there are other influential deep uncertainties. To account for the impact of deep uncertainties and to simplify the discussion, we design four scenarios. Our ``standard'' scenario assumes a fossil fuel resource base consistent with the best guess resource estimates from \citet{mohrProjectionWorldFossil2015} and uses a truncated normal prior distribution for the half-saturation year of zero-carbon technologies (that is, the year when that technology achieves a 50\% share of the energy mix). This distribution assigns a 2.5\% prior probability to half-saturation between 2020 and 2050, and has its mode at 2100. Our ``low fossil-fuel'' and ``high fossil-fuel'' scenarios use the same prior distribution over the zero-carbon technology half-saturation year, but adopt the low and high resources estimates, respectively, from \citet{mohrProjectionWorldFossil2015}. These three scenarios allow us to explore the implications of varying assumptions about fossil fuel supplies, with the associated knock-on effects for energy-generating costs, on CO$_2$ emissions. 

We also examine the sensitivity of our projections to a more pessimistic set of prior beliefs about zero-carbon technology penetration. In this ``delayed zero-carbon'' scenario, we assign only a 2.5\% prior probability to global half-saturation by 2100. To isolate the impact of this prior distribution, we use the same fossil fuel resource constraint as in the standard scenario.

In each of these scenarios, if a set of parameters results in fossil-fuel consumption exceeding the size of the resource base for any fuel, it is excluded from the calibration set. This induces a trade-off between economic growth and fossil-fuel consumption, such that economic growth is slowed down if the economy is still dependent on limited fossil-fuel resources. While we do not explicitly model the prices of fossil fuels, this resulting effect is analogous to the effect of increasing prices due to relative resource scarcity. However, for a fixed rate of economic growth, technology succession can occur earlier than the latest year associated with the fossil fuel constraint, representing changing demand as the driver of technological change, rather than restrictions in supply. 

By itself, this fossil-fuel constraint does not rule out ahistorical fuel substitution dynamics despite the statistical calibration procedure, as uncertainties in and correlations between the emissions intensities of the fossil-fuel technologies, the half-saturation years of the various technologies, and the rate of technological penetration create large degrees of freedom in mapping historical economic growth to CO$_2$ emissions. For example, some parameter sets might imply a 50\% zero-carbon share and/or a less than 10\% coal share in 2020 (see Fig. S1 in Online Resource 1). To constrain this behavior, we add a further constraint on technology shares in 2019. Based on fuel share data from the last ten years~\citep{BpStatisticalReview2020}, we require that the coal share in 2019 is between 20\% and 30\% and that the zero-carbon share is between 10\% and 20\%. These windows were chosen to be relatively wide to acknowledge year-to-year volatility in these shares. This constraint captures one of the most notable differences between our ``baseline'' scenarios and the no-policy ``baseline'' SSP-RCP scenarios, as the share of these various technologies has been influenced by past and current policies such as subsidies and tax credits.

Our model simulations also include error terms to account for discrepancies between the data and the model outputs~\citep{brynjarsdottirLearningPhysicalParameters2014}. These error terms are generated conditionally using the VAR(1) discrepancy process (see Section S3 of Online Resource 1) on the discrepancies over a period of years which are selected to be fixed. For example, the size of the discrepancy terms for projections are made conditionally on the model-data discrepancies over the 1820--2019 historical period to preserve the calibrated variance structure of the discrepancies. 

\section{Calibration Results}

\begin{figure}
    \centering
    \includegraphics{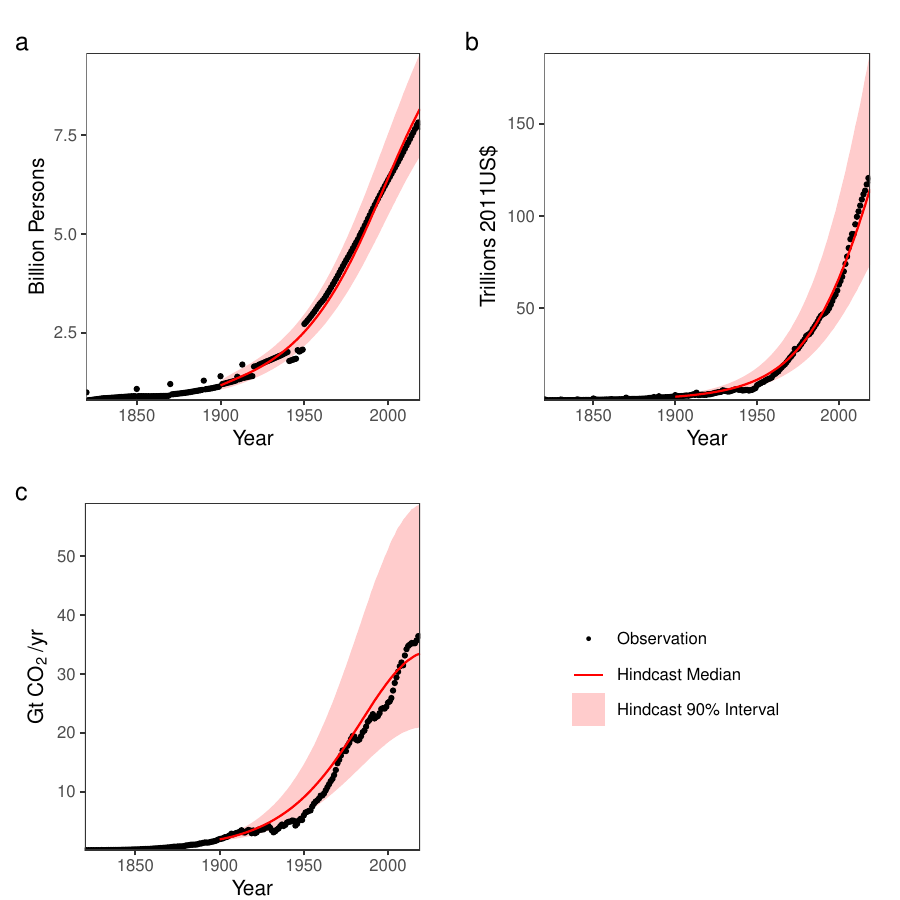}
    \caption{\textbf{Model hindcast results across all model outputs --} Median and 90\% credible intervals for model hindcasts from 1900--2019 conditioned on data from 1820--1899. Hindcasts are provided for all three model components: (a) global population, (b) gross world product, and (c) global CO$_2$ emissions.}
    \label{fig:hindcast}
\end{figure}

We conduct a hindcasting exercise and perform cross-validation to assess potential biases of the calibrated model and to evaluate out-of-sample projection skill. For the hindcast, we make projections for the period from 1900--2019 conditional on data from 1820--1899 (see Fig.~\ref{fig:hindcast}). The median growth simulations slightly underestimate post-2000 economic growth. As a result, the model was unable to reproduce the rapid growth in emissions starting around 2000, while it did capture the recent slowing of the emissions growth rate. The hindcasts also demonstrate a large degree of underconfidence in the post-1950 period.

We further test the model predictive skill use a $k$-fold cross-validation procedure. We randomly sample fifty hold-out test data sets (each corresponding to forty years). The model is re-calibrated with the remaining training data, and we generate simulated data for the held-out years conditional on the training data. The average cross-validation coverage of the 90\% credible intervals for the held-out data are 92\% for population and economic output and 93\% for emissions, suggesting good out-of-sample predictive performance despite the hindcast's underconfidence. The underconfidence seen in the hindcast is likely the result of the accumulation of errors over the 120-year hindcasting period. These errors can grow quite large, as the model's projections and the data do not diverge strongly until later in the 20th century across all parameter vectors in the calibrated parameter set. However, we note that this level of underconfidence is similar to that seen in \citet{rafteryLessWarming21002017} despite the many differences between the two models, suggesting that this may be a consequence of using a statistically calibrated, relatively simple model.

\section{Projections of Future CO$_2$ Emissions}

\begin{figure}
    \centering
    \includegraphics[width=.9\textwidth]{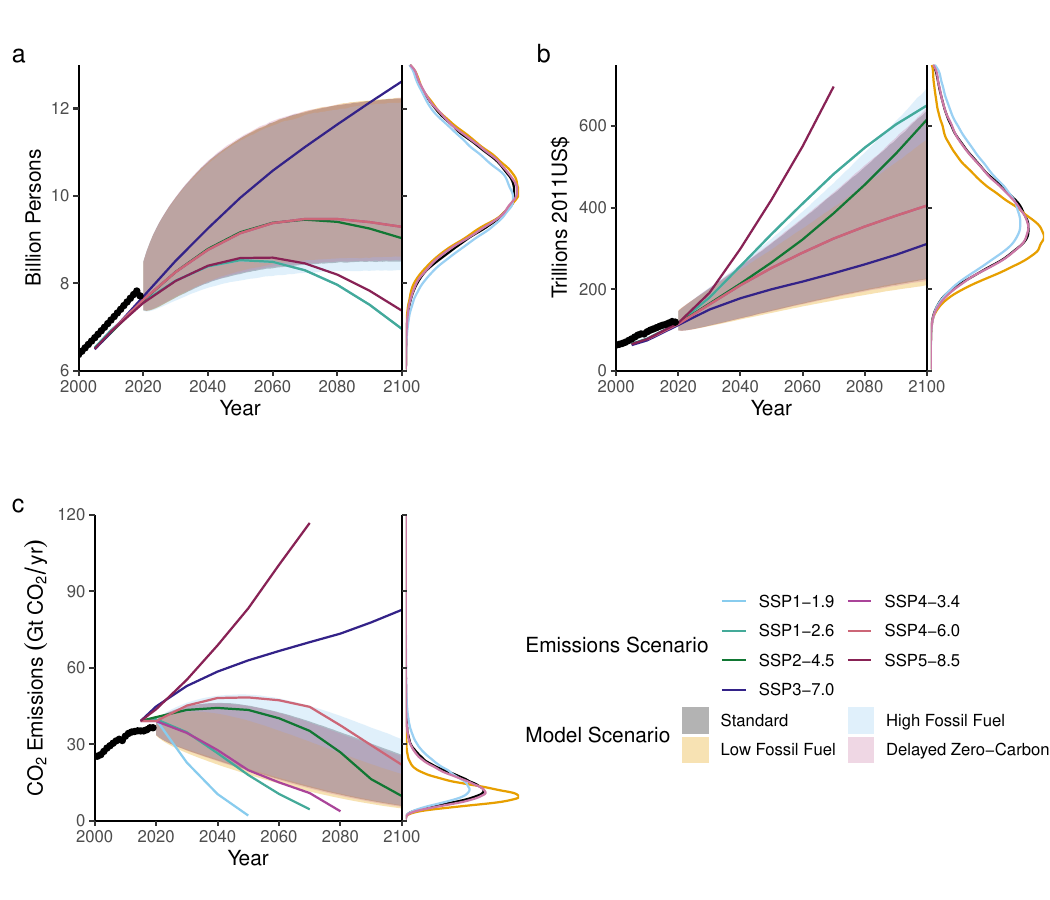}
    \caption{\textbf{Projections of model outputs ---} Projections of (a) global population (billions of persons), (b) gross world product (trillions 2011US\$), and (c) annual carbon dioxide emissions projections from 2020--2100 for the considered scenarios. The shaded regions are the central 90\% credible intervals. Black dots represent  observations from 2000--2019. The marginal distributions of each output in 2100 are shown on the right. The relevant quantities from the marker baseline SSP-RCP emissions scenarios~\citep{riahiSharedSocioeconomicPathways2017, rogeljMitigationPathwaysCompatible2018, oneillScenarioModelIntercomparison2016} are shown for comparison by the lines.}
    \label{fig:emissions}
\end{figure}

Deep uncertainty about fossil fuel resources causes more variability in CO$_2$ projections than prior beliefs about zero-carbon penetration (Fig.~\ref{fig:emissions}c). Under the base scenario assumptions, the median cumulative emissions from 2020--2100 are 2200 GtCO$_2$, with the 90\% prediction interval covering 1500--3000 GtCO$_2$. The lower end of the prediction interval, as well as the median levels, are only slightly influenced by the fossil fuel constraint, varying by 100-200 GtCO$_2$ at most, which is expected as the constraint does not affect simulations with faster decarbonization than is required by the supply-side limit. On the other hand, the supply constraint matters more at the upper end, as the low-fossil fuel scenario prediction interval's upper bound is 2600 GtCO$_2$, with the high-fossil fuel interval extending to 3400 GtCO$_2$. This is largely the result of varying the coal constraint, as residual coal emissions can get close to the resource limit when the penetration of lower- and zero-emitting technologies is slower and economic growth rates are more rapid. This is consistent with the analysis by \citet{capellan-perezLikelihoodClimateChange2016}, conducted using the more detailed GCAM, which found that, when considering resource constraints, cumulative CO$_2$ emissions was mainly sensitive to the size of the coal resource base. Both of these results further agree with the argument by \citet{ritchie1000GtCCoal2017} and \citet{ritchieWhyClimateChange2017} that the relative plausibility of higher-emissions scenarios is dependent on their assumptions about coal resource availability and utilization.

The main impact of the technology penetration constraint is to sharply constrain the half-saturation years of the various generating technologies ($\tau_i$ in Fig. S2), and, due to correlations, the inferred emissions intensity of the higher-emissions technology ($\rho_2$ in Fig. S2). Due to this sharpening of the half-saturation year of the zero-emissions technology ($\tau_4$ in Fig. S2), there is little difference between the standard and delayed zero-carbon scenario penetrations, despite the influence of the CO$_2$ emissions expert assessment from \citet{hoNotAllCarbon2019} in the calibration (see Fig. S3 and Fig. S4; including the CO$_2$ expert assessment increases the upper tail areas of the non-low fossil fuel scenarios, but does not result in a separation of the standard and delayed zero-carbon scenarios despite their different priors, as the marginal distributions of the emissions parameters are the same regardless of what assessments are used). The impact of the constraint on emissions projections can be seen in Fig. S5. The technological constraint also induces a typical half-saturation year in the second-half of the 21st century, with the 90\% central credible interval between 2057 and 2086, and a median half-saturation year of 2071 (see Fig. S6 for how the distribution of shares of our technologies change over the remainder of the century). These quantities are relatively insensitive to the resource constraint. 

After accounting for the additional emissions from 2015--2019, the base median estimate of 2100 GtCO$_2$ is slightly higher than that from the ``Continued'' forecast from \citet{liuCountrybasedRateEmissions2021}, which had a median cumulative emissions from 2015--2100 of 2100 GtCO$_2$. It should be noted that this \citet{liuCountrybasedRateEmissions2021} scenario assumes that nations will meet their Nationally Determined Contributions (NDCs), and therefore assumes additional mitigation policies beyond those captured by our model's data. The relatively small difference is likely due to the logistic penetration curve in our model, which causes a faster rate of decarbonization after half-saturation compared to the constant rate of post-NDC carbon intensity decrease in the \citet{liuCountrybasedRateEmissions2021} scenario. However, all of our projections are much lower than the base forecasts by \citet{liuCountrybasedRateEmissions2021}, which could be due to their spatial disaggregation, as economic growth with slower decarbonization can result in lower cumulative emissions from countries and regions with relatively small economies and, therefore, low levels of emissions despite high projected utilizations of fossil fuels. In our model, the global economy decarbonizes uniformly, the rate of which (due to the penetration constraint) is heavily influenced by the energy mix in developed countries. This suggests that our projections may be biased towards lower levels of emissions by the level of aggregation, as globally-aggregated projections can mask the importance of regional deviations from the global trend~\citep{pretisCarbonDioxideEmissionintensity2017}.

Our model's projections are broadly consistent with the extrapolations from current International Energy Agency (IEA) projections made by \citet{hausfatherEmissionsBusinessUsual2020}. Indeed, in the IEA Stated Policies scenario~\citep{ieaWorldEnergyOutlook2020}, emissions in 2030 are 36 GtCO$_2$/yr, which is the median value of our high fossil fuel scenario (90\% predictive interval of 28--46 GtCO$_2$/yr) and close to the median (35 GtCO$_2$/yr in 2030) of our standard scenario's projections (90\% predictive interval 28--45 GtCO$_2$/yr) and low-fossil fuel projections (median 34 GtCO$_2$/yr, 90\% predictive interval 27--43 GtCO$_2$/yr).  The IEA's Sustainable Development scenario, which projects 27 GtCO$_2$/yr, is just in the lower tail of most of our projections. Unsurprisingly, then, the bulk of our predictive distributions, as shown in Fig.~\ref{fig:emissions}, correspond to the range deemed by \citet{hausfatherEmissionsBusinessUsual2020} to be ``likely'' given current policies. However, as with the \citet{hausfatherEmissionsBusinessUsual2020} analysis, our model does not account for land-use emissions, which could increase emissions sufficiently to be more consistent with higher-emitting SSP-RCP scenarios~\citep{schwalmRCP8TracksCumulative2020}. Further, our model cannot by its structure account for the possibility of technological backsliding, which would result in increase in global emissions intensities due to \emph{e.g.} an increase in coal use in varying regions around the world or the replacement of nuclear plants with fossil-fueled generation. These possible trend breaks are also not reflected in projections such as those from the IEA, which are mainly based on extrapolating current trends~\citep{skeaOutlooksExplorationsNormative2021}. For example, as noted by \citet{skeaOutlooksExplorationsNormative2021}, this class of projections anticipates nuclear generation growth by 39\% on average.

\begin{figure}
    \centering
    \includegraphics[width=.5\textwidth]{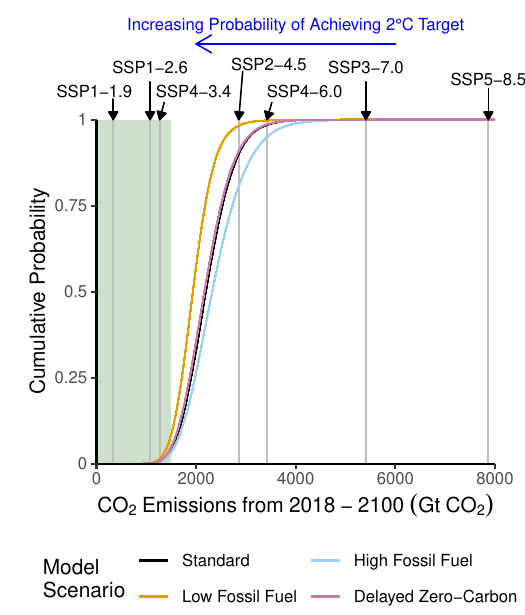}
    \caption{\textbf{Cumulative emissions projections from 2018-2100 --} Cumulative density functions for cumulative emissions projections for the four model scenarios. The grey lines are the cumulative emissions over this period for the labelled IPCC scenario. The green region represents cumulative emissions which are consistent with at least a 50\% probability of achieving the 2$^\circ$C Paris Accords target~\citep{rogeljMitigationPathwaysCompatible2018}.}
    \label{fig:cdf}
\end{figure}

The projections for the remainder of the century broadly span the area captured by those SSP-RCP scenarios which include additional decarbonization, but no NET penetration (see Figs.~\ref{fig:emissions} and~\ref{fig:cdf}. In particular, SSP2-4.5 closely tracks the upper range of the standard and delayed zero-carbon scenarios until about 2070, when decarbonization occurs more rapidly in that SSP-RCP scenario than in our projections. SSP4-6.0, on the other hand, is a tail scenario under these assumptions, but characterizes the upper level of the likely range of the high-fossil fuel scenario. The no-policy SSP5-8.5 and SSP3-7.0 scenarios are well outside of our distributions, but for different reasons. Our model considers SSP5-8.5 implausible due to the runaway levels of economic growth (see Fig.~\ref{fig:emissions}b) fueled by the wide utilization of extremely large coal reserves. On the other hand, SSP3-7.0 has slower economic growth, but both large population growth (Fig.~\ref{fig:emissions}a) and continued fossil fuel use.

The impact of projected economic growth is crucial for interpreting these projections as well-calibrated or biased-downward. Due to our choice of calibration period, our projected rates of per-capita economic growth from 2018--2100 are lower than 2\% per year (see Fig. S7). This is much lower than the expert assessment reported in \citet{christensenUncertaintyForecastsLongrun2018}, which was used in both this analysis and \citet{gillinghamModelingUncertaintyIntegrated2018a}. With the projected levels of growth from our calibrated model and the rate of decarbonization induced by the penetration constraint, almost all of our simulations show the rate of carbon intensity decreasing more rapidly than the increase in per-capita economic growth. The difference between these two Kaya identity components is greater than in the forecast by \citet{liuCountrybasedRateEmissions2021}, which could be the result of our spatial aggregation, as discussed earlier.

Our base scenario simulations contain a small fraction (4\%) of outcomes where emissions are reduced rapidly enough to be consistent with at least a 50\% probability of achieving the 2$^\circ$C Paris Agreement target\citep{rogeljMitigationPathwaysCompatible2018}, though we note that this is only in reference to CO$_2$ emissions, not the CO$_2$ equivalent of emissions of other greenhouse gases, and that we also neglect the impact of emissions from land-use change. We classify these simulations using a classification tree~\citep{breimanClassificationRegressionTrees1984, therneauRpartRecursivePartitioning2019}. These states of the world are characterized by combinations of the rate of technological penetration and the emissions intensity of the lower-emitting fossil technology, rather than the half-saturation year of the zero-carbon technology. In particular, this outcomes only occurs in the simulations if the emissions intensity of the lower-emitting fossil technology is below the 44th percentile of the marginal posterior (see Fig. S2 and S8; $\rho_3$). If the rate of technological penetration is sufficiently slow (~4\%/year, which is the 88th percentile of the marginal posterior; see Fig. S2 and S8; $\kappa$), the emissions intensity of the lower-emitting fossil technology must be no greater than its 32nd percentile (approximately 0.044 GtCO$_2$/US\$2011). On the other hand, there are no such combinations which characterize relatively high-emissions outcomes, which is likely due to the complex interactions driving rapid economic growth as well as the impact at the high end of residual emissions from coal utilization and the resource constraint.

\begin{figure}
\centering
\includegraphics{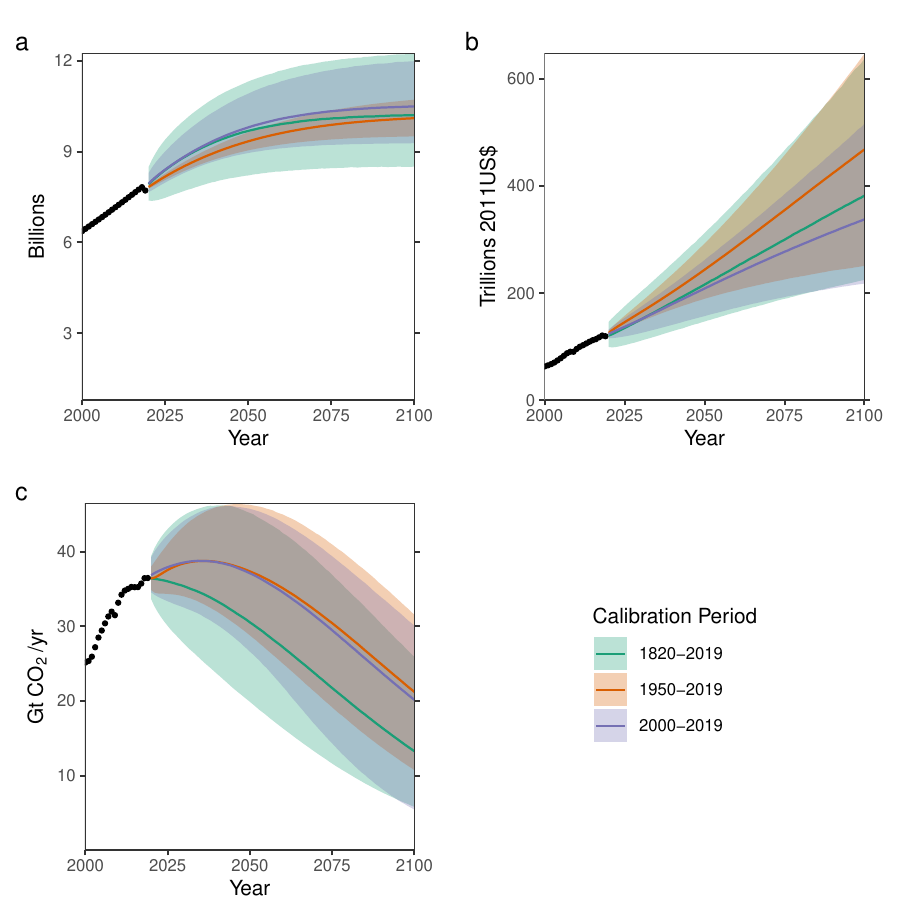}
\caption{\textbf{Model projections using different calibration periods --} Model projections based on calibrations using data from 1820--2019, 1950--2910, and 2000--2019 for (a) global population, (b) gross world product, and (c) global CO$_2$ emissions. For each calibration period, the line is the median of the projections, and the ribbon is the 90\% credible interval. These projections were made using the ``standard'' scenario assumptions about fossil fuel resource limits and zero-carbon technology half-saturation year priors.}
\label{fig:hindcast-length}
\end{figure}

These results depend on our choice of century-scale data (Fig.~\ref{fig:hindcast-length}).  Using data from 1950 results in higher projected economic growth and emissions by the end of the century despite a reduced overall level of uncertainty due to the growth in both economic output and production after World War II and the exclusion of previous boom-bust cycles, recessions, and depressions. On the other hand, the high-economic growth tail from the 1820--2019 and 1950--2019 calibrations is not present in the 2000--2019 calibration due to the exclusion of the economic growth second half of the 20th century. This has interesting implications, as even the other calibrations did not result in particularly rapid projected economic growth for the remainder of the 21st century. The use of century-scale data also results in lower emissions projections than the use of data starting in the 20th century due to differences across all of the emissions parameters (see Fig. S7). Using century-scale calibration data results in a higher inferred technological penetration rate, as well as more constrained half-saturation estimates for the fossil technologies. The end result is that the use of shorter data sets produces projections in which emissions peak later in the century, resulting in higher emissions even when economic growth is slower.

\section{Cumulative Emissions Sensitivities}

\begin{figure}
    \centering
    \includegraphics[width=.7\textwidth]{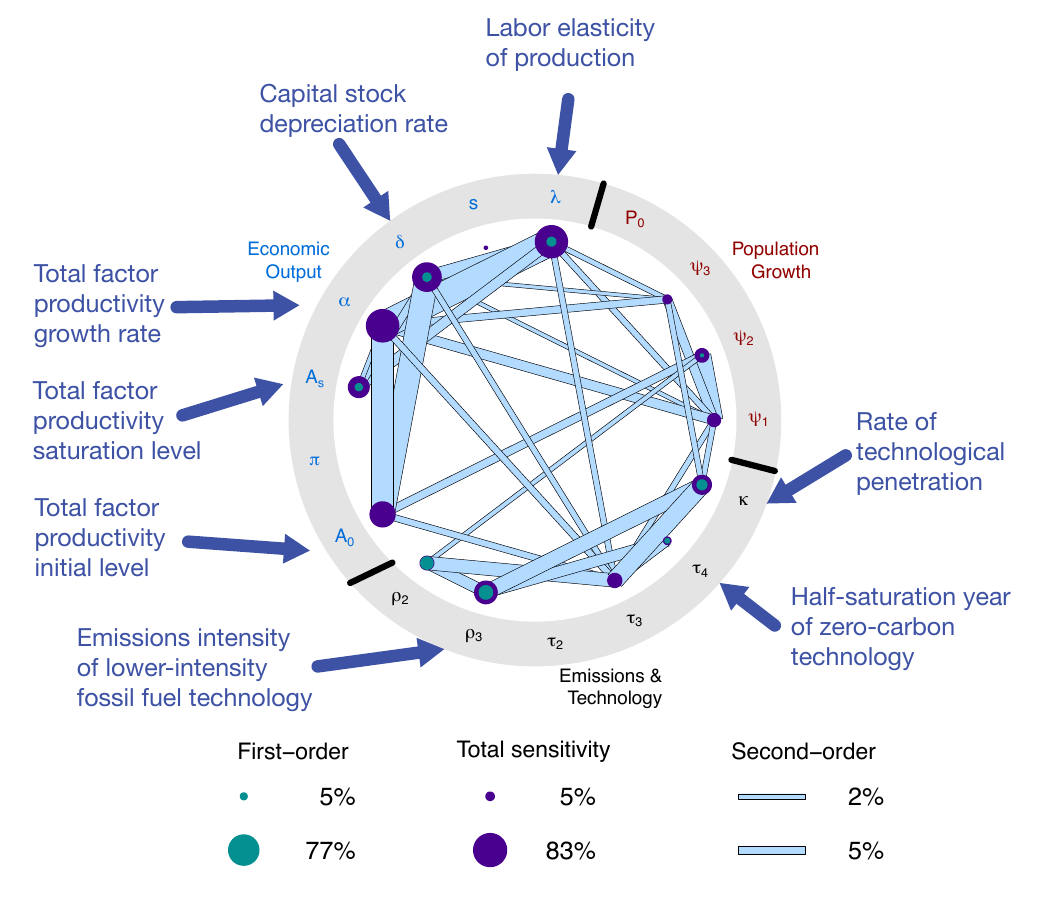}
    \caption{\textbf{Global sensitivities of cumulative emissions to model variables ---} Global sensitivity~\citep{sobolSensitivityEstimatesNonlinear1993, sobolGlobalSensitivityIndices2001} indices for the decomposition of variance of cumulative emissions under our standard scenario from 2018--2100. The computation of sensitivity index estimates is described in the Section S4 of Online Resource 1. Filled green nodes represent first-order sensitivity indices, filled purple nodes represent total-order sensitivity indices, and filled blue bars represent second-order sensitivity indices for the interactions between parameter pairs. Important parameters are labelled with their role in the model. Other model variable names are defined in Table S1. Sensitivity values exceeding thresholds are provided in Tables S4 and S5.}
    \label{fig:sensitivity}
\end{figure}

Cumulative emissions variability is driven mainly by uncertainties in interacting economic and technology dynamics, with a much smaller contribution from population dynamics~(Fig. \ref{fig:sensitivity}). Cumulative emissions exhibit statistically-significant sensitivities (in the sense that the 95\% confidence interval of the sensitivity index does not contain zero) to all model parameters other than the initial population ($P_0$), the half-saturation year of the more-intensive fossil fuel technology ($\tau_2$), and the labor force participation rate ($\pi$). This illustrates a challenge of constructing parsimonious models for projecting emissions. The first- and higher-order sensitivities to a large number of parameters illustrates the complexity of the system dynamics, even for this highly aggregated, relatively simple IAM. Uncertainties in several economic variables, including those characterizing total factor productivity growth, explain a large fraction of the variability in cumulative emissions, showing the importance of improving our understanding of economic growth dynamics in addition to technological shifts in the energy sector if we are to further constrain future emissions projections.

Economic variables matter more through higher-order sensitivities and interactions with other parameters, while variables related to emissions intensities and technological substitutions have a more direct influence. This is due to the translation of economic growth into emissions through the mixture of emitting technologies within the model. While it may be surprising that the half-saturation year of the zero-carbon technology ($\tau_4$) has a relatively small influence, this results from the same model dynamics discussed earlier when characterizing low-emissions tail scenarios. As the marginal posterior of $\tau_4$ is relatively narrow and is limited to the second-half of the 21st century, the rate of technological penetration ($\kappa$, which influences the shape of the logistic curves) and the emissions intensities of the fossil technologies ($\rho_2$ and $\rho_3$) explain most of the uncertainty in mapping economic output to emissions. These large sensitivities highlight the need for updated accounting of emissions factors to help constrain and update projections of CO$_2$ emissions. The emissions intensity then combines with economic growth, which is dominated by total factor productivity, or TFP (as discussed by \citet{nordhausQuestionBalanceWeighing2008} and \citet{gillinghamModelingUncertaintyIntegrated2018a}, as well as the dynamics of capital, including depreciation (controlled by the capital depreciation rate $\delta$ and the capital elasticity of production ($1-\lambda$, where $\lambda$ is the labor elasticity).

It is worth discussing further the importance of TFP growth in this context. The growth rate ($\alpha$) explains a large degree of cumulative emissions variability through its intersection with other parameters, even though it has a non-statistically sensitivity first-order sensitivity. In this model, there is a strong interaction between the TFP growth rate and the elasticity of production with respect to capital, as these directly affect production growth. Future TFP growth dynamics are likely to be nonstationary compared to our inferences from the historical record, due to the increasing penetration of automation in the global economy, though a more detailed representation of these dynamics might also include trend breaks in TFP growth and an accounting of the displacement effect of automation on the labor share~\citep{acemogluAutomationNewTasks2019}. Our model's lack of ability to account for accelerating TFP growth, and therefore even faster economic growth, could partially explain our lower growth projections compared to \citet{christensenUncertaintyForecastsLongrun2018}, as well as the lack of higher emissions scenarios in our projections.

We do not observe large sensitivities to population dynamics. This is consistent with the analyses from \citet{rafteryLessWarming21002017}, which found only a 2\% sensitivity of CO$_2$ emissions in 2100 to population, as well as \citet{gillinghamModelingUncertaintyIntegrated2018a}, which found a 10\% sensitivity. This could be the result of our model's global aggregation, as some regionally-focused analyses have found that population growth is the Kaya identity component most typically associated with changes in CO$_2$ emissions~\citep[\emph{e.g.}][]{vanruijvenBaselineProjectionsLatin2016}. However, our analysis does contradict the finding of \citet{vanvuurenConditionalProbabilisticEstimates2008}, which identitifed population as a major driver of CO$_2$ emissions.

\section{Discussion}
In this analysis, we produced baseline probabilistic CO$_2$ projections from a simple, mechanistically-motivated integrated assessment model calibrated on century-scale historical data under several realizations of different deep uncertainties. This type of modeling exercise has several potential virtues. Our model runs rapidly enough to be statistically calibrated using a long data set as well as to be subjected to a global sensitivity analysis. The coupled uncertainty- and sensitivity- analyses allow us to identify important linkages across different modeling components and interpretable model parameters, illustrating the complexity of the joint social-economic-technical system which ultimately results in CO$_2$ emissions. Some of these linkages would not be directly seen when using a more computationally expensive model which might preclude the required number of model evaluations. By comparing the resulting probabilities across the different scenarios corresponding to the considered deep uncertainty, we could explore the impacts of varying assumptions. For example, we could see the impact of fossil fuel resource uncertainty on the high-emissions upper tail. 

One clear lesson is that even for this relatively stylized and highly-aggregated model, calibration using historical data is not itself sufficient to fully constrain the model dynamics. For one, the choice of calibration period matters substantially in projecting economic growth and emissions intensities. Second, even this simple model has enough degrees of freedom to produce inaccurate energy-generating technology shares without strong constraints (which still result in underconfident hindcasts). The imposition of these constraints results in consistency with IEA Stated Policy projections~\citep{ieaWorldEnergyOutlook2020} through 2030. However, our projections assume that current technological substitution trends will continue or accelerate, and do not account for the possibility of technological backsliding. It would be possible and interesting to use a modeling framework similar to ours to understand the extent to which backsliding and increased economic growth could combine to produce high-emissions outcomes which seem unlikely based on current and historical trends. Our projections are also made under the baseline assumption that no new mitigation policies will be implemented. This assumption is unlikely to be true in practice, particularly as the impacts of climate change become more apparent, and we use it purely as a counterfactual.

Global aggregation also runs the risk of masking local and regional dynamics which could be influential in determining how future emissions change, particularly in regions which have so far either no contributed much to total CO$_2$ emissions or which are experiencing rapid economic growth~\citep{pretisCarbonDioxideEmissionintensity2017}. These could be strong enough to cause trend breaks from the historical dynamics, resulting in emissions higher than those projected by our modeling exercise or the IEA, particularly if currently planned mitigation policies are not fully implemented or are abandoned in the face of political or economic pressures. With its current formulation, our model, and therefore the projections, are incapable of addressing this possibility, which could result in the type of technological backsliding discussed earlier. Our model could, however, be adapted to explore scenarios which allow for non-constant rates of technological penetration or the possibility of an older emitting technology to recapture higher shares of the global energy mix. Another example is the potential impact of population growth in regions such as sub-Saharan Africa, which were a cause of the relative lack of sensitivity of emissions to population growth in \citet{rafteryLessWarming21002017}. It would be interesting to use a spatially-disaggregated version of our model to explore the combinations of population and economic growth and technological penetrations which would result in increased emissions outside of the likely range reported here.

There are several other caveats that are important to mention. While the economic outputs in our analysis suggest that the higher end of economic growth forecasts, such as those elicited in \citet{christensenUncertaintyForecastsLongrun2018}, will not be achieved (even before accounting for shocks such as the COVID-19 pandemic), this is partially dependent on both our choice of calibration period as well as the structure of our economic model, which rules out the possibility of trend breaks in growth resulting from \emph{e.g.} automation. Additionally, our study is silent on the impacts of negative emissions technologies. Changes in climate policies could result in these technologies becoming viable and widespread prior to the end of the century. This would shift emissions downward starting from the point when these technologies begin to penetration. This effect depends, however, on the currently-uncertain details of these technologies. One extension of this study might be to include these deep uncertainties, producing projections which are conditional on both the negative emissions technologies and the rate of penetration of a sample technology. Finally, while baseline scenarios are a useful counterfactual, climate policies evolve, and the odds of policies remaining the same through 2100 are nearly zero. The resulting changes in incentives for technology and energy-use would necessarily alter the dynamics captured by our model calibration.

It is important to stress that our analysis has also neglected the large effect of the Earth-system response to changes in CO$_2$\citep{friedlingsteinUncertaintiesCMIP5Climate2014, boothNarrowingRangeFuture2017, quilcailleUncertaintyProjectedClimate2018a}. From the perspective of managing climate risk, emissions projections and forecasts should be understand in the context of these large climate-system uncertainties. It would be unwise to ignore the implications of higher-emissions scenarios for risk analysis by focusing only on their emissions trajectories, as higher levels of radiative forcing and changes to global mean temperature could be obtained from lower emissions levels than those used in scenario generation. Moreover, these scenarios have important value in climate modeling analyses due to their high signal-to-noise ratio. Our analysis here is not intended to downplay the value of these scenarios for understanding the climate system or for climate risk analysis. We also did not model non-CO$_2$ greenhouse gas emissions.

Ultimately, even with these caveats, simple, mechanistically-motivated models have a role to play in understanding the uncertainties associated with future climate risks. Many of the limitations discussed above could be addressed by transparently creating or modifying such a model and comparing projections, parameter distributions, and sensitivities across assumptions. The systems which produce anthropogenic CO$_2$ are sufficiently complex that the ability to map out which parts of the system interact with other parts to produce higher- or lower emissions is important to improve our understanding of climate risk. More computationally-expensive and complex models are valuable in representing these complex system dynamics from the ground up, but may not be amenable to this type of analysis without the use of emulators, which smooth out the model dynamics to some degree and may restrict the number of parameters which can be considered (depending on the type of emulator). Ultimately, all models are wrong~\citep{boxScienceStatistics1976}, but the use of multiple models with varying levels of complexity and which transparently make different assumptions can help us gain a holistic view of future climate risk and its drivers.

%\begin{acknowledgements}
\section*{Acknowledgements}
The authors thank Louise Miltich and Alexander Robinson for their inputs and contributions. We thank Arnulf Grübler, Jonathan Koomey, Robert Lempert, Michael Obersteiner, Brian O’Neill, Hugh Pitcher, Steve Smith, Rob Socolow, Dan Ricciuto, Mort Webster, Xihao Deng, Tony Wong, Jonathan Lamontagne, Emily Ho, Wei Peng, Casey Helgeson, and Billy Pizer for discussions and comments. 
%\end{acknowledgements}

\section*{Declarations}
\subsection*{Funding}
This work was partially supported by the National Science Foundation (NSF) through the Network for Sustainable Climate Risk Management (SCRiM) under NSF cooperative agreement GEO-1240507 and the Penn State Center for Climate Risk Management. Any opinions, findings, and conclusions or recommendations expressed in this material are those of the authors and do not necessarily reflect the views of the funding entities.

\subsection*{Conflicts of interest/Competing interests}
The authors are aware of no conflicts of interest.

\subsection*{Availability of data and material}
All data used in this study is available at https://github.com/vsrikrish/IAMUQ.

\subsection*{Code availability}
All code used in this study is available at https://github.com/vsrikrish/IAMUQ.

\subsection*{Authors' contributions}
V.S., R.T., and K.K. designed the research. V.S. and K.K. conducted the research. V.S., Y.G., and K.K. analyzed the results. V.S., Y.G., R.T., and K.K. wrote the paper.

\subsection*{Ethics approval}
Not applicable

\subsection*{Consent to Participate}
Not applicable

\subsection*{Consent for publication}
Not applicable

\bibliography{main}
\end{document}

\maketitle

\captionsetup{labelformat=stab}

\section{Model Structure}

Model outputs are globally-aggregated population (in billions), gross world product (in trillions 2011 US\$), and carbon emissions (in GtC/yr), with an annual temporal resolution. The model generates these outputs using a set of three coupled modules: (1) population; (2) economic output; (3) carbon emissions. We could have followed the Kaya Identity and included energy use. However, CO$_2$ emissions are not measured but instead imputed from energy use. Including energy use would not add information, but it would increase the dimensionality of the model. 

There is a bidirectional coupling between population and economic output: population affects labor inputs through the labor force participation rate, while per-capita income affects the rate of population growth. CO$_2$ emissions are a consequence of economic output through a mixture of emitting technologies with varying emissions intensities.

\subsubsection*{Population}

We model population growth using a logistic model~\citep{Cohen1995-xh}, modified by an income-sensitive growth rate. At time $t$, population $P_t$ is given by
\begin{equation*}P_t = P_{t-1}[1 + \psi_1(y_{t-1} / (\psi_2 + y_{t-1}))((\psi_3 - P_{t-1})/\psi_3)],\end{equation*}
where $y$ is annual per-capita income, $\psi_1$ is the annual population growth rate, $\psi_2$ is the half-saturation parameters with respect to per-capita consumption, and $\psi_3$ is the logistic carrying capacity. This model structure allows for interactions between per-capita income and population growth. Note that this equation implies that the population grows before stabilizing near saturation; other projections have been made which assume population peaks and then decreases~\citep{Lutz2001-iz}.

\subsection*{Economic Output}

To model gross world product, we use a Cobb-Douglas production function in a Solow-Swan model of economic growth. Total world production $Q_t$ at time $t$ is
\begin{equation*}Q_t = A_t L_t^\lambda K_t^{1-\lambda},\end{equation*}
where $A$ is total factor productivity, $L$ is labor input, $K$ is capital input, and $\lambda$ is the elasticity of production with respect to labor. Each year, total production is divided between consumption and investment,
\begin{equation*}Q_t = U_t + I_t = (1-s)Q_t + sQ_t,\end{equation*}
where $s$ is the savings rate, which we assume to be constant.

Growth in total factor productivity (TFP) occurs exogenously. The dynamics of long-term technological change are deeply uncertain\citep{Starr1973-qc, Ausubel1995-mp}. Following~\citet{Nordhaus1983-eo}, we allow TFP to saturate as the population ages and becomes less innovative:
\begin{equation*}A_t = A_{t-1} + \alpha A_{t-1}\left[1 - \left(\frac{A_{t-1}}{A_s}\right)\right],\end{equation*}
where $\alpha$ is the TFP growth rate and $A_s$ is the TFP saturation level.

Capital stock growth occurs through a balance of depreciation and investment from the previous time step,
\begin{equation*}K_t = (1-\delta)K_{t-1} + I_{t-1},\end{equation*}
where $\delta$ is the capital depreciation rate, which is constrained to be less than the savings rate $s$. Labor input is determined by
\begin{equation*}L_t = \pi P_t,\end{equation*}
where $\pi$ is the labor force participation rate. Labor is initialized using the (uncertain) initial population $P_0$, while capital is initialized using the steady-state relationship
\begin{equation*}K_0 = \left(\frac{sA_0}{\delta}\right)^{1/\lambda}L_0.\end{equation*}

\subsection*{Carbon Emissions}

We model the link between economic output and anthropogenic CO$_2$ emissions from fossil fuel burning and cement production using a time-dependent carbon intensity of production. That is, we assume that energy is a derived demand, rather than a factor of production. The carbon emissions $C_t$ at time $t$ are
\begin{equation*}C_t = Q_t \phi_t,\end{equation*}
where $\phi$ is the carbon intensity. $\phi$ is modeled as a weighted average of four broadly defined technologies,
\begin{equation*}\phi_t = \sum_{i=1}^4 \gamma_{i,t}\rho_i,\end{equation*}
where $\gamma_{i,t}$ is the fraction of the economy invested in technology $i$, which has a technology-specific carbon intensity $\rho_i$. This assumes that each fuel is used uniformly across end-uses. We set the carbon intensity of technology 1 to zero to represent pre-industrial economic activity, which had negligible fossil fuel emissions (while human activity did produce CO$_2$ emissions during this period~\citep{Ruddiman2003-pm}, the primary driver of these emissions was land-use change, which we do not consider). The carbon intensities of the technologies 2 and 3 are estimated from observations with the constraint $\rho_2 \geq \rho_3$. This constraint represents the transition from a higher carbon intensity technology, analogous to coal, to a lower carbon intensity technology, analogous to oil and natural gas.  We set the carbon intensity of technology 4 to zero to simulate the penetration of low-carbon technologies such as nuclear and renewables. The time dynamics of $\gamma_i$ are approximated as logistic penetration curves,
\begin{align*}
\gamma_{1,t} &= 1 - \frac{1}{1 + \exp(-\kappa(t-\tau_2))} \\
\gamma_{2,t} &= \frac{1}{1 + \exp(-\kappa(t-\tau_2))} - \frac{1}{1 + \exp(-\kappa(t-\tau_3))}\\
\gamma_{3,t} &= \frac{1}{1 + \exp(-\kappa(t-\tau_3))} - \frac{1}{1 + \exp(-\kappa(t-\tau_4))}\\
\gamma_{4,t} &= 1 - \frac{1}{1 + \exp(-\kappa(t-\tau_4))},
\end{align*}
where $\kappa$ is the rate at which technologies penetrate and $\tau_i$ is the time at which technology $i$ has penetrated half the market. This type of logistic penetration model can reasonably approximate observed energy substitution dynamics~\citep{Marchetti1977-dw, Grubler1999-qc}.

\section{Model Calibration}

We use Bayesian data assimilation to calibrate the model and estimate parametric and predictive uncertainty. Bayesian statistical methods allow for the fusion of data with prior information via Bayes' theorem~\citep{Bayes1763-at}, yielding probabilistic parameter estimates and projections.  As our model outputs are necessarily greater than zero, we construct our likelihood function using log-scale residuals. The residuals between the model estimates and the observations are modeled using a vector autoregression process of order 1 (VAR(1)). We use this specification due to the auto- and cross-correlations of the residuals when an independent and identically distributed error assumption is made. The lag-1 auto-correlations are 0.60 for population, 0.49 for gross world product, and 0.94 for emissions. The lag-0 cross-correlations range from 0.13 to 0.58. The mathematical specification of the residual structure model and derivation of the corresponding likelihood function is provided in the Section S3.

The fossil fuel resource constraint is used in the evaluation of the likelihood. After running the model with a proposed set of parameters, the output emissions are compared to the constraint. If the total emissions violate the constraint, the likelihood of the parameters is assigned a value of zero. Similarly, if the technology shares of the global economy are outside of the specified penetration constraint windows, the likelihood of the parameters is assigned a likelihood of zero.

We use Markov chain Monte Carlo (MCMC) with the Metropolis-Hastings algorithm~\citep{Metropolis1953-rv, Hastings1970-zq} to draw samples from the posterior distribution for each modeling scenario. Four chains are run to facilitate convergence checks. Each chain is initialized at a maximum a posteriori (MAP) estimate for that scenario and is run for two million iterations. A burn-in period of 500,000 iterations is discarded at the start of each chain. Both the chain and burn-in lengths are determined using a combination of visual inspection and the Gelman-Rubin diagnostic~\citep{Gelman1992-da}. Plots of the resulting marginal prior and posterior distributions for the standard scenario are provided in Fig. S8. The pairs plot of the posterior distribution is provided in Fig. S10.

We show how alternate priors for important parameters, described in Table S3, affect the resulting projections in Fig. S11. Posterior distributions for the standard scenario corresponding to the inclusion of the three expert assessments (including none and all) are shown in Fig. S4. The impact of the expert assessments on projections is shown in Fig. S3.

\section{Derivation of VAR(1) Likelihood Function}

Let $\mathbf{z}_t = (z_{1t}, z_{2t}, \ldots, z_{mt})^T$ and $\mathbf{M}(\bm{\theta})_t = (M(\bm{\theta})_{1t}, M(\bm{\theta})_{2t}, \ldots, M(\bm{\theta})_{mt})^T$ be vectors of the observations and model outputs (corresponding to model parameters $\bm{\theta})$ at time $t$, respectively. In our case, $m=3$, corresponding to population, economic output, and CO$_2$ emissions. We model the difference between $\mathbf{z}_t$ and $\mathbf{M}(\bm{\theta})_t$ with a vector autoregressive (VAR) component $\mathbf{x}_t$, which allows model errors to be correlated over time. 

The model for the observations has the form
\begin{align*}
    \mathbf{z}_t &= \mathbf{M}(\theta)_t + \mathbf{x}_t + \bm{\varepsilon}_t \\
    \mathbf{x}_t &= A\mathbf{x}_{t-1} + \mathbf{w}_t,
\end{align*}
where $\mathbf{w}_t \sim N(\mathbf{0}, W)$ and $\bm{\varepsilon}_t \sim N(\mathbf{0}, D)$, with D a diagonal matrix of observation errors. The vectors $\mathbf{x}_t$, $\mathbf{w}_t$, and $\bm{\varepsilon}_t$ have dimension $m$, defined similarly as $\mathbf{z}_t$ and $\mathbf{M}(\bm{\theta})_t$. We assume that the observation errors are independent over time with known errors $D$, and $\mathbf{w}_t$ are the white noise process of the VAR model. Hence, $\text{Cov}(\bm{\varepsilon}_t, \bm{\varepsilon}_s) = 0$ and $\text{Cov}(\mathbf{w}_t, \mathbf{w}_s) = 0$. Further, we assume that the process $\mathbf{x}_t$ is weakly stationary with $E(\mathbf{x}_t) = 0$. 

The marginal covariance of $\mathbf{x}_t$ can be derived from solving the following for $\Sigma_x$:
\begin{align*}
    \Sigma_x &= \text{Cov}(\mathbf{x}_t) \\
    &= E[(A\mathbf{x}_{t-1} + \mathbf{w}_t)(A\mathbf{x}_{t-1} + \mathbf{w}_t)'] \\
    &= A E(\mathbf{x}_{t-1} \mathbf{x}_{t-1}')A' + E(\mathbf{w}_t \mathbf{w}_t') \\
    &= A \Sigma_x A' + W,
\end{align*}
which gives
\begin{align*}
    \text{vec}(\Sigma_x) &= (A \otimes A) \text{vec}(\Sigma_x) + \text{vec}(W)\\
    \text{vec}(\Sigma_x) &= (I - A \otimes A)^{-1} \text{vec}(W).
\end{align*}

Conditional on the model outputs, $\mathbf{z}_t \sim N(\mathbf{M}(\theta)_t, \Sigma_z)$, where 
\begin{align*}
    \Sigma_z &= \text{Cov}(\mathbf{z}_t) \\
    &= E[(\mathbf{x}_t + \bm{\varepsilon}_t)(\mathbf{x}_t + \bm{\varepsilon}_t)'] \\
    &= E(\mathbf{x}_t \mathbf{x}_t') + E(\bm{\varepsilon}_t \bm{\varepsilon}_t') \\
    &= \Sigma_x + D.
\end{align*}

The covariance for any two observations with lag $h$ is
\begin{align*}
    \text{Cov}(\mathbf{z}_t, \mathbf{z}_{t-h}) &= A^h \text{Cov}(\mathbf{x}_{t-1}, \mathbf{x}_{t-h}) \\
    &= A^h \Sigma_x.
\end{align*}
Hence, the likelihood function for observations $\mathbf{z} = (\mathbf{z}_1', \ldots, \mathbf{z}_T')'$ is 
\begin{equation*}
    L(\mathbf{z}; \bm{\theta}, A, D, W) = |\Sigma|^{-1/2}\exp\left(-\frac{1}{2}\left(\mathbf{z} - M(\bm{\theta})\right)'\Sigma^{-1}\left(\mathbf{z}-M(\bm{\theta})\right)\right),
\end{equation*}
with $\mathbf{M}(\bm{\theta}) = (\mathbf{M}(\bm{\theta})_1', \ldots, \mathbf{M}(\bm{\theta})_T')'$ and
\begin{equation*}
    \Sigma = \begin{pmatrix}\Sigma_x + D & (A\Sigma_x)' & \ldots & (A^{n-1}\Sigma_x)' \\
    A\Sigma_x & \Sigma_x + D & \ldots & (A^{n-2}\Sigma_x)' \\
    \vdots & \vdots & \ddots & \vdots \\
    A^{n-1}\Sigma_x  & A^{n-2}\Sigma_x & \ldots & \Sigma_x + D\end{pmatrix}.
\end{equation*}

\newpage

\section{Prior Distributions and Sensitivities}

\subsection*{Default Priors} 

We complete the specifications of the Bayesian hierarchical model by assigning prior distributions to all unknown parameters. Supplemental Tables 1 and 2 list the prior distributions for these parameters. Prior distributions are specified by their family (\emph{e.g.} normal or uniform) and a lower and upper bound. Normal-family distributions (including normal and log-normal distributions) are specified using their 2.5\% and 97.5\% central probability limits, while uniform distributions are specified using their lower and upper bounds.

\newgeometry{margin=0.25in}
\begin{table}
\centering
\begin{threeparttable}[b]
\setlength\extrarowheight{2pt}
\caption{Prior distributions used in calibration for the model parameters. Lower and upper bounds are absolute bounds for uniform distributions and central 95\% probability intervals for normal and log-normal distributions.}
\begin{tabularx}{\textwidth}{SBBSSSB}
\toprule
Parameter & Description & Units & Prior & Lower Bound & Upper Bound  & Reference \\
\midrule
$\psi_1$ & population growth rate & 1/year & normal &  0.0001 & 0.15 & this study \\
$\psi_2$ & half-saturation constant & 1000\$/(year capita) & uniform & 0 & 50 & this study \\
$\psi_3$ & population carrying capacity & billions & normal & 6.9 & 14.4 & \citet{Lutz1997-sx}; \tnote{a}\\ 
$P_0$ & population in 1700 & billions & normal & 0.3 & 0.9 & \citet{Maddison2003-iu}; \tnote{b}\\
$\lambda$ & elasticity of production with respect to labor & dimensionless & normal & 0.6 & 0.8 & \citet{Romer2012-gw}\\
$s$ & savings rate & dimensionless & normal & 0.22 & 0.26 & this study\tnote{c}\\
$\delta$ & capital depreciation rate & 1/year & uniform & 0.01 & 0.14 & \citet{Nordhaus1994-fd, Nadiri1996-vc}\\
$\alpha$ & total factor productivity growth rate & 1/year & normal & 0.0007 & 0.0212 & this study\\
$A_s$ & saturation level of total factor productivity & dimensionless & uniform & 5.3 & 16.11 & \citet{Nordhaus1994-fd};\tnote{d}\\
$\pi$ & labor participation rate & dimensionless & normal & 0.62 & 0.66 & this study;\tnote{e}\\
$A_0$ & total factor productivity in 1700 & dimensionless & uniform & 0 & 3 & this study;\tnote{f} \\
$\rho_2$ & carbon intensity of technology 2 & kg carbon/2011US\$ & normal & 0 & 0.75 & this study\\
$\rho_3$ & carbon intensity of technology 3 & kg carbon/2011US\$ & normal & 0 & 0.75 & this study\\
$\tau_2$ & half-saturation year of technology 2 & year & uniform & 1700 & 2100 & this study \\
$\tau_3$ & half-saturation year of technology 3 & year & uniform & 1700 & 2100 & this study \\
$\tau_4$ & half-saturation year of technology 4 & year & normal\tnote{g} & 2050 & 2150 & this study \\
$\kappa$ & rate of technological penetration & 1/year & uniform & 0.005 & 0.2 & \citet{Grubler1991-ka}\\
\bottomrule
\end{tabularx}
\begin{tablenotes}
\item [a] The lower bound is the peak population in the 2.5\% scenario; the upper bound the 2100 population in the 97.5\% scenario.
\item [b] The lower bound is the minimum of the four alternative estimates of \citet[Table B-1]{Maddison2003-iu} minus the standard deviation; the upper bound is the maximum plus the standard deviation.
\item [c] The global average gross savings rate between 1977 and 2017 is 24\% with a standard deviation of 1.1\% \citep{World_Bank2018-aq}. The range given here is consistent with that distribution.
\item [d] We use the ratio of the 2005 level to the long-term saturation level with a uniform probability density function of $\pm 50\%$.
\item [e] The global average labor force participation rate between 1990 and 2018 is 64\% with a standard deviation of 1.4\% \citep{World_Bank2019-sf}. The range given here is consistent with that distribution. 
\item [f] The best guess is obtained using $A_0 = Q_0{\lambda/(1+\lambda)}(\delta/s)^{\lambda/(1+\lambda)}(\pi P_0)^{\lambda^2/(1+\lambda)}$.
\item [g] This distribution was truncated from below at 2020.
\end{tablenotes}
\end{threeparttable}
\label{tab:model-prior}
\end{table}

\restoregeometry

\begin{table}[h!]
\centering
\begin{threeparttable}[b]
\setlength\extrarowheight{2pt}
\caption{Prior distributions used in calibration for the statistical parameters.}
\begin{tabularx}{\textwidth}{BBBSSSB}
\toprule
Parameter & Description & Units & Prior & Lower Bound & Upper Bound  & Reference \\
\midrule
$a_{ii}$ & diagonal entries of VAR coefficient matrix $A$ & dimensionless & normal & 0 & 1 & this study \\
$a_{ij}, i \neq j$ & off-diagonal entries of VAR coefficient matrix $A$ & dimensionless & normal & -1 & 1 & this study \\
$\sigma_1$ & log-scale population innovation variance &  (log-billions)$^2$ & log-normal & 0 & $\infty$ & this study\tnote{a}\\
$\sigma_2$ & log-scale GWP innovation variance  &  (log-trillions 2011USD\$)$^2$ & log-normal & 0 & $\infty$ & this study\tnote{a}\\
$\sigma_3$ & log-scale emissions innovation variance  &  (log-GtC/yr)$^2$ & log-normal & 0 & $\infty$ & this study\tnote{a}\\
$\varepsilon_1$ & log-scale population observation error variance & (log-billions)$^2$ & log-normal & 0 & $\infty$ & this study\tnote{a}\\
$\varepsilon_2$ & log-scale GWP observation error variance & (log-trillions 2011USD\$)$^2$ & log-normal & 0 & $\infty$ & this study\tnote{a}\\
$\varepsilon_1$ & log-scale emissions observation error variance & (log-GtC/yr)$^2$ & log-normal & 0 & $\infty$ & this study\tnote{a}\\
\bottomrule
\end{tabularx}
\begin{tablenotes}
\item [a] Log-normal distributions have log-scale means of -1 and log-scale standard deviations of 1.
\end{tablenotes}
\end{threeparttable}
\label{tab:stat-prior}
\end{table}

\subsection*{Sensitivity to Priors}

Table S3 specifies alternate priors for particular parameters to detect the sensitivity of projections to the choice of priors. These parameters were selected because they were either identified as important by the sensitivity analysis (Fig. 5) or they were not updated by the Bayesian inversion (Fig. S8). The alternate prior specifications were selected to include unbounded prior ranges (when the prior was previously uniform) or fatter tails (when the prior was previously normal). Other priors are kept the same from the standard scenario.

Fig. S10 shows the projections resulting from calibrating the model with these prior distributions. In general, the projections are identical, and the qualitative features of the marginal distribution in 2100 are preserved.

\begin{table}[h!]
\centering
\begin{threeparttable}[b]
\setlength\extrarowheight{2pt}
\caption{Prior distributions used in calibration for the model parameters. Lower and upper bounds are absolute bounds for uniform distributions and central 95\% probability limits for normal and log-normal distributions.}
\begin{tabularx}{\textwidth}{SBBSSSB}
\toprule
Parameter & Description & Units & Prior & Lower Bound & Upper Bound \\
\midrule
$\lambda$ & elasticity of production with respect to labor & dimensionless & log-normal & 0.6 & 0.8 \\
$s$ & savings rate & dimensionless & log-normal & 0.22 & 0.26 \\
$A_s$ & saturation level of total factor productivity & dimensionless & normal & 5.3 & 16.11\\
$\pi$ & labor participation rate & dimensionless & log-normal & 0.62 & 0.66\\
\bottomrule
\end{tabularx}
\end{threeparttable}
\label{tab:sens-prior}
\end{table}

Figure S2 shows the projections resulting from alternate prior distributions for $\tau_4$, the half-saturation year of the zero-carbon technology. The ``alternate'' prior is a normal distribution (truncated from below at 2020) with a 95\% central confidence interval from 2050 to 2250. We also show projections from the delayed zero-carbon prior from the main manuscript.

\section{Sobol' Sensitivity Indices}
To compute the Sobol' sensitivity indices~\citep{Sobol1993-vi, Sobol2001-xh}, the  model  parameter  space  is  sampled  using  quasi-random  Saltelli  sampling~\citep{Saltelli2010-yw}. To compute the second-order interactive sensitivity indices, $M = 2n(d+1)$ samples must be generated,  where $d$ is the number of uncertain parameters and $n$ is sufficiently large to achieve the desired level of precision.  The larger the value of $n$, the more precise the estimates of the sensitivity indices, but the procedure will be more computationally expensive. We use $n=1e5$ samples. Confidence intervals for the estimates are computed using a bootstrapping analysis with $1e4$ replicates. We evaluate the convergence of the estimates based on a manual inspection of the confidence interval width. Tables S4 and S5 report statistically significant indices that were above selected thresholds.

\begin{table}
    \centering
    \caption{First- and total-order Sobol' Sensitivity indices for each sampled parameter that accounts for greater than 1\% of total variability in cumulative emissions from 2018-2100.  The 95\% confidence interval of each index is provided in parentheses. Only statistically significant variables greater than 1\% are reported. Total-order sensitivities can add up to be greater than 1 due to multiple second-order interactions.}
    \begin{tabular}{r|c|c|c}
    Parameter & Description & Total-Order Index & First-Order Index \\
    \midrule
    $\psi_1$ & population growth rate & 0.15 (0.15, 0.16) & -- \\
    $\psi_2$ & half-saturation constant & 0.17 (0.17, 0.17) & -- \\
    $\psi_3$ & population carrying capacity & 0.08 (0.08, 0.08) & -- \\ 
    $\lambda$ & elasticity of production with respect to labor & 0.86 (0.86, 0.87) & 0.08 (0.07, 0.08)\\
    $s$ & savings rate & 0.02 (0.01, 0.02) & -- \\
    $\delta$ & capital depreciation rate & 0.68 (0.68, 0.69) & 0.07 (0.07, 0.07) \\
    $\alpha$ & total factor productivity growth rate & 0.87 (0.87, 0.88) & -- \\
    $A_s$ & saturation level of total factor productivity & 0.37 (0.37, 0.38) & 0.05 (0.05, 0.06)  \\
    $A_0$ & total factor productivity in 1700 & 0.54 (0.53, 0.54) & -- \\
    $\rho_2$ & carbon intensity of technology 2 & 0.17 (0.17, 0.17) & 0.14 (0.14, 0.15) \\
    $\rho_3$ & carbon intensity of technology 3 & 0.44 (0.44, 0.44) & 0.17 (0.16, 0.17) \\
    $\tau_3$ & half-saturation year of technology 3 & 0.18 (0.18, 0.18) & -- \\
    $\tau_4$ & half-saturation year of technology 4 & 0.06 (0.06, 0.06) & 0.03 (0.02, 0.03) \\
    $\kappa$ & rate of technological penetration & 0.31 (0.31, 0.31) & 0.10 (0.09, 0.10)
    \end{tabular}
    \label{tab:soboltot}
\end{table}

\begin{table}
    \centering
    \caption{Second-order Sobol' Sensitivity indices for each interaction between sampled parameters that accounts for greater than 10\% of total variability in cumulative emissions from 2018-2100.  The 95\% confidence interval of each index is provided in parentheses. Only statistically significant interactions greater than 10\% are reported.}
    \begin{tabular}{r|c|c}
    Parameter 1 & Parameter 2 & Second-Order Index \\
    \midrule
    $\psi_1$ & $\psi_2$ & 0.25 (0.24, 0.25)\\
    $\lambda$ & $\delta$ & 0.13 (0.12, 0.13) \\
    $\lambda$ & $\alpha$ & 0.58 (0.57, 0.59) \\
    $\delta$ & $A_0$ & 0.43 (0.43, 0.44) \\
    $\alpha$ & $A_0$ & 0.38 (0.38, 0.38) \\
    $\rho_2$ & $\tau_3$ & 0.14 (0.14, 0.15) \\
    $\rho_3$ & $\tau_4$ & 0.10 (0.09, 0.11) \\
    $\rho_3$ & $\kappa$ & 0.17 (0.17, 0.18) \\
    $\tau_3$ & $\kappa$ & 0.25 (0.24, 0.26)
    \end{tabular}
    \label{tab:sobolsec}
\end{table}

\clearpage
\bibliography{supplement}

\clearpage
\section*{Supplemental Figures}
\captionsetup{labelformat=sfig}

\begin{figure}[hb!]
    \centering
    \includegraphics{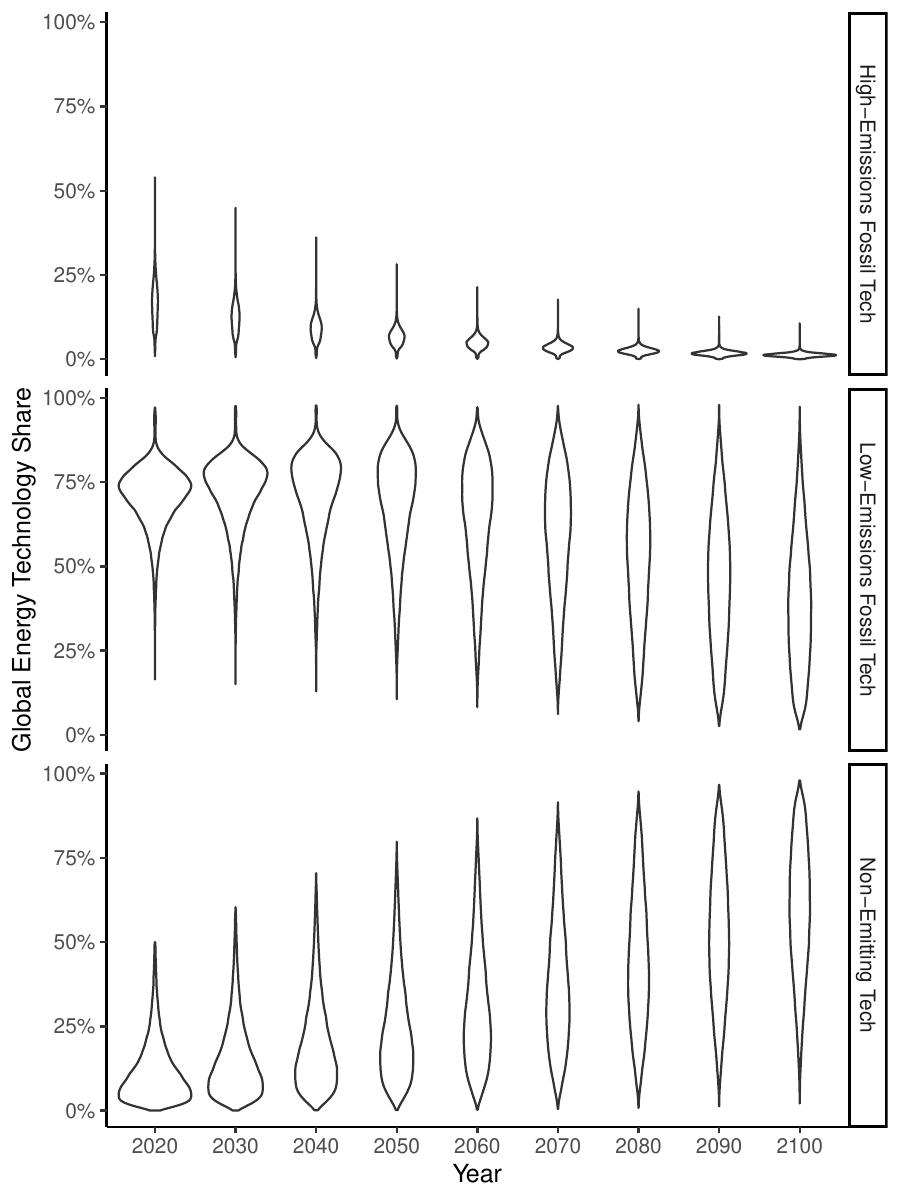}
    \caption{\textbf{Technology shares without a penetration constraint --} Violin plots of the shares of the various emitting technologies in our model from 2020--2100 without an active technology penetration constraint in 2019.}
    \label{fig:tech-share-nopen}
\end{figure}

\begin{figure}
    \centering
    \includegraphics[width=\textwidth]{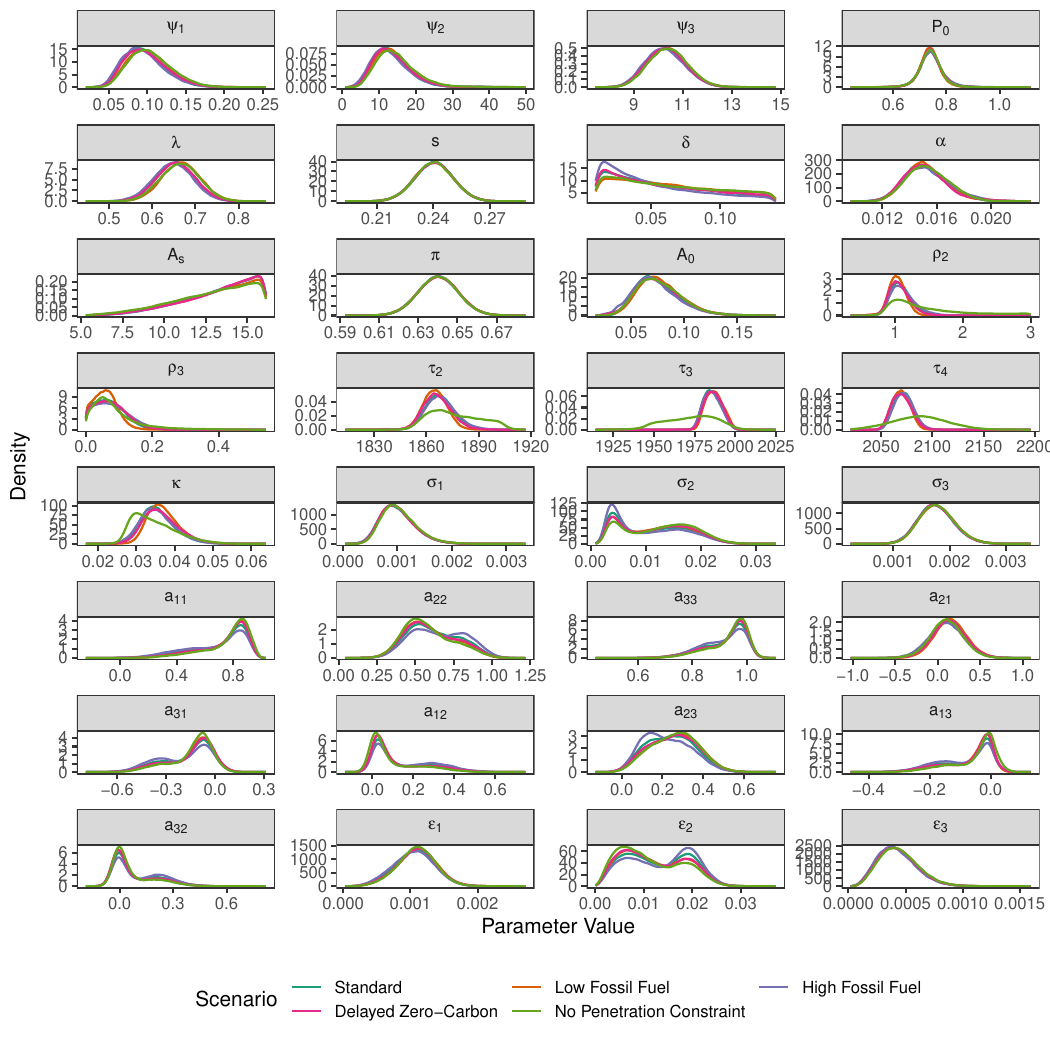}
    \caption{\textbf{Parameter posterior distributions by model scenario --} Marginal posterior distributions for the model parameters under each model scenario.}
    \label{fig:dist-scenario}
\end{figure}

\begin{figure}
    \centering
    \includegraphics[width=\textwidth]{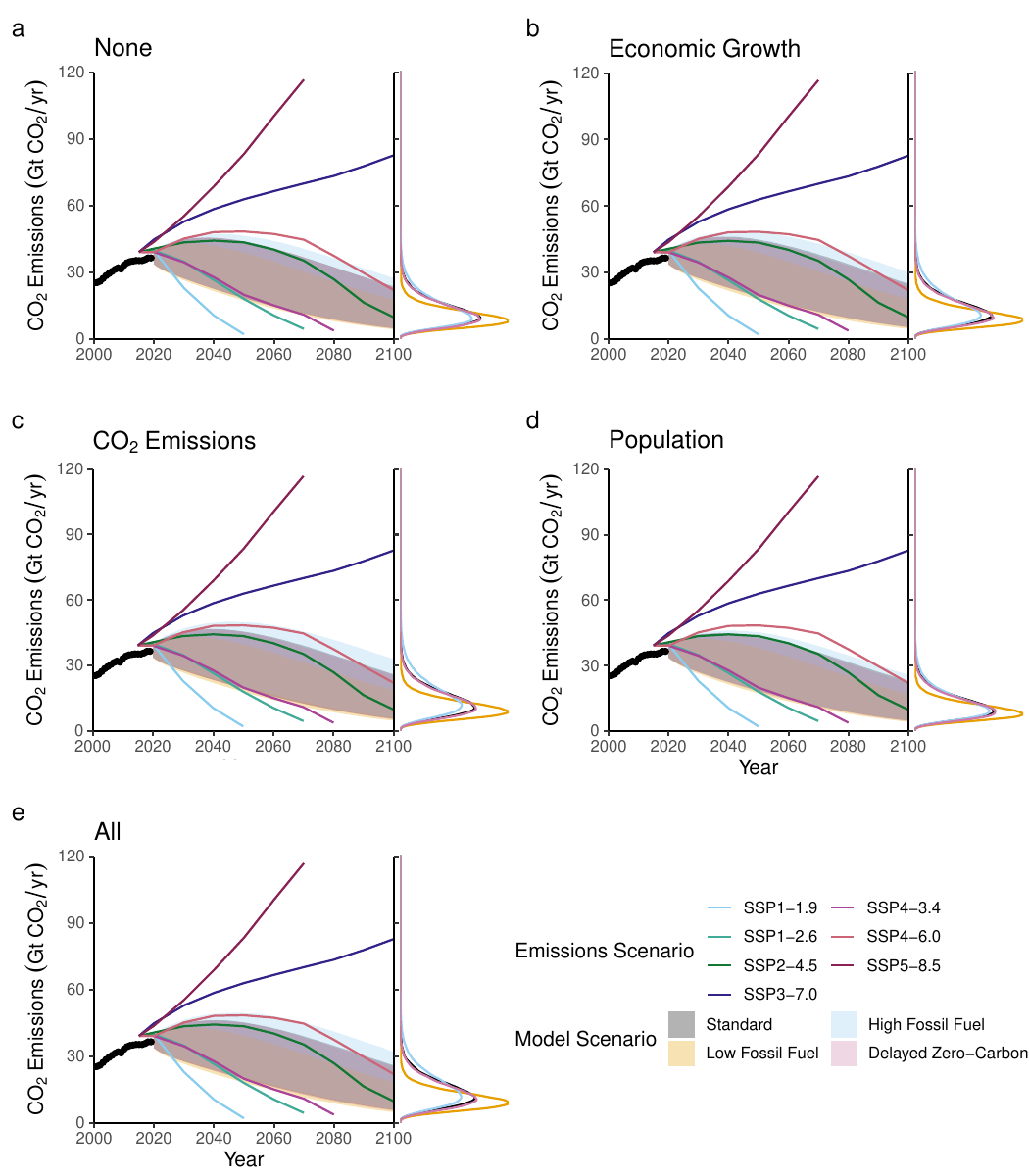}
    \caption{\textbf{Sensitivity of carbon dioxide emissions projections to expert assessments --} Time series and marginal distribution in 2100 for annual CO$_2$ emissions projections with respect to varying expert assessment assimilations: a) No expert assessments; b) only Christensen et al (2018)$^21$; c) only Ho et al (2019)$^8$; d) only United Nations (2019); e) all three.}
    \label{fig:co2-expert}
\end{figure}

\begin{figure}
    \centering
    \includegraphics[width=\textwidth]{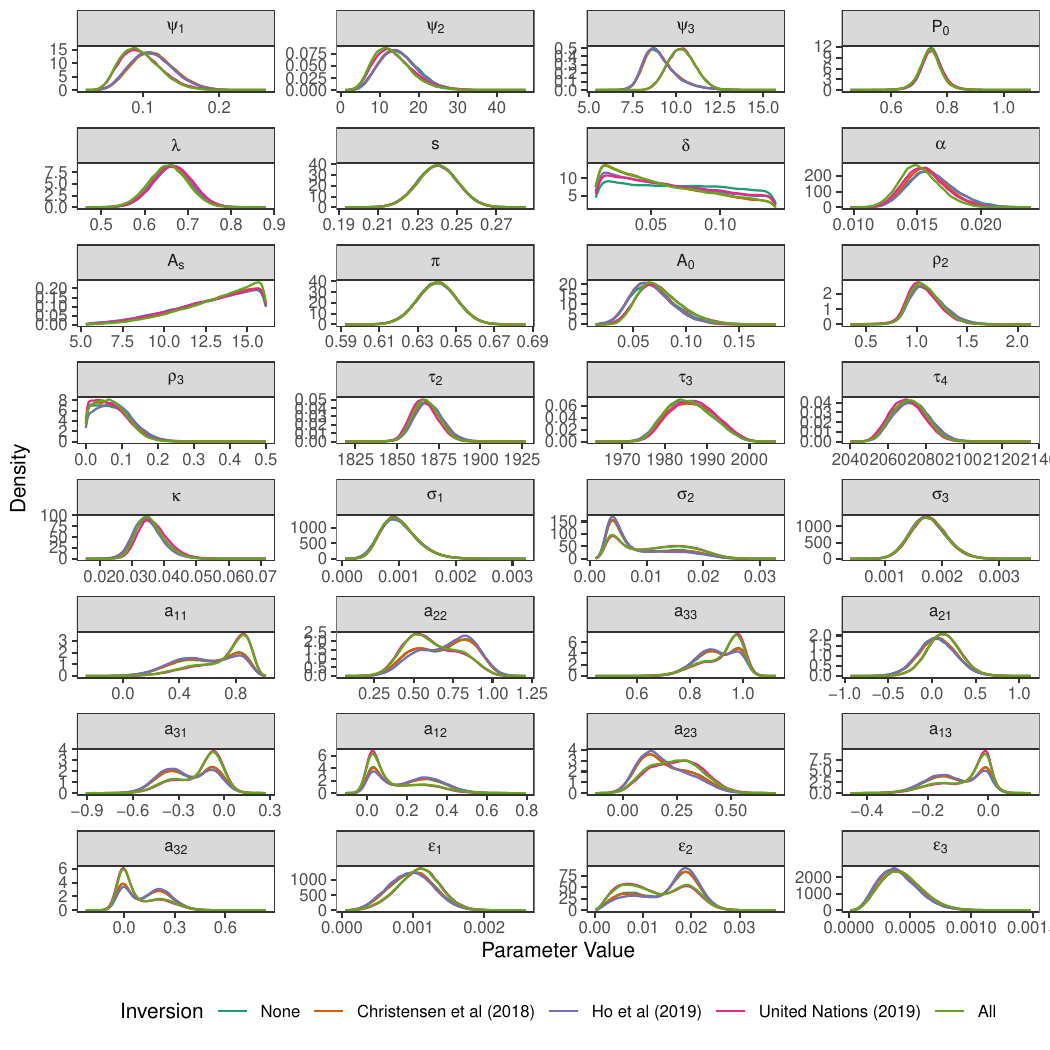}
    \caption{\textbf{Expert assessment impact on model parameter posterior distributions --} Posterior distributions for the model parameters under the standard scenario with respect to calibrations assimilating different expert assessments.}
    \label{fig:dist-expert}
\end{figure}

\begin{figure}
    \centering
    \includegraphics{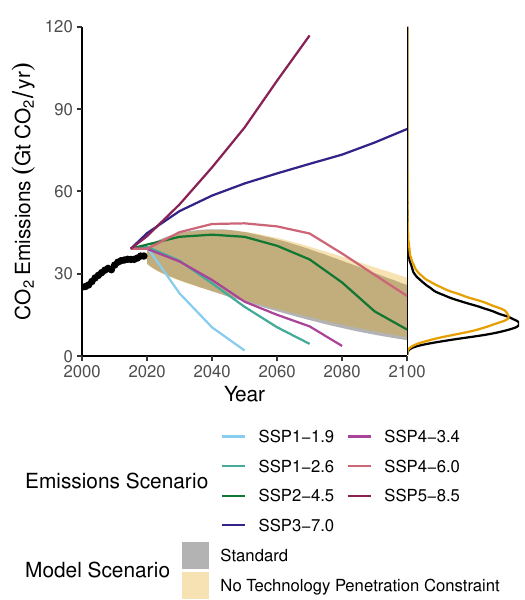}
    \caption{\textbf{Impact of technology penetration constraint on projected emissions --} Time series and marginal distribution in 2100 for annual CO$_2$ emissions under the standard scenario assumptions, with and without the technology penetration constraint.}
    \label{fig:co2-techpen}
\end{figure}

\begin{figure}[hb!]
    \centering
    \includegraphics{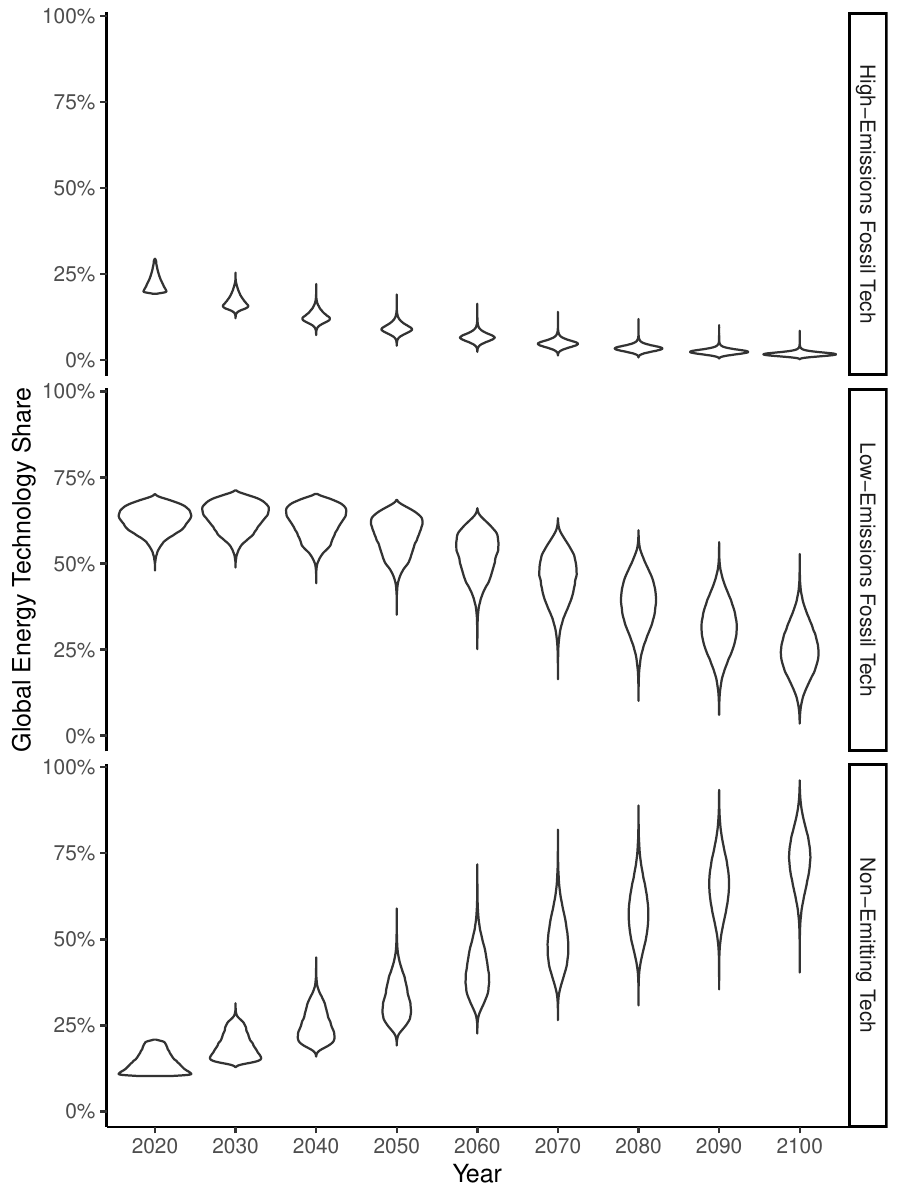}
    \caption{\textbf{Technology shares with the penetration constraint --} Violin plots of the shares of the various emitting technologies in our model from 2020--2100 with our technological constraint.}
    \label{fig:tech-share-base}
\end{figure}

\begin{figure}
    \centering
    \includegraphics{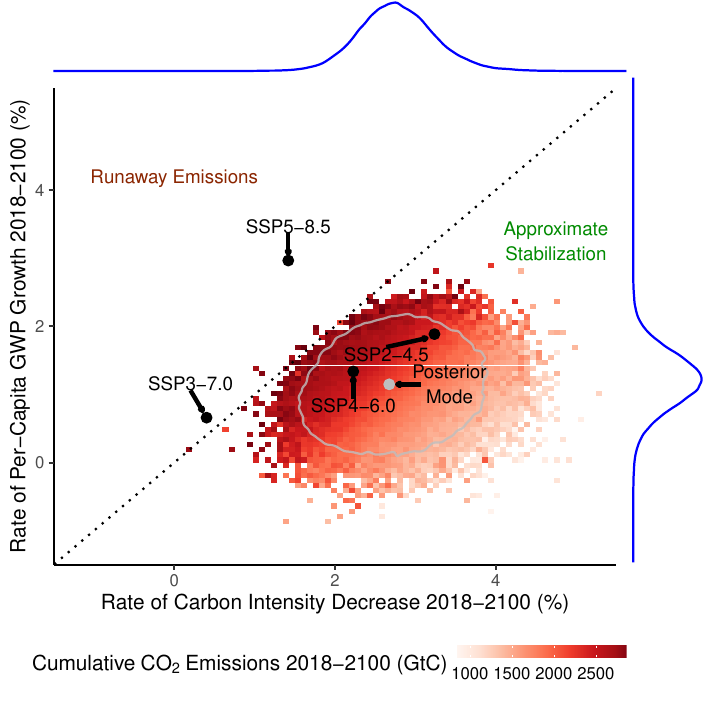}
    \caption{\textbf{Cumulative emissions from 2018-2100 by rates of global economic growth and carbon intensity decrease --} Mean cumulative emissions from 2018-2100 are shown with respect to each binned region of average annual rates of global economic growth and carbon intensity decrease. Relevant SSP-RCP scenarios are shown as black dots. The grey contour is the 95\% posterior region and the posterior mode is represented by a grey dot. The dotted diagonal line is the 1-1 line corresponding to equal rates of economic growth and carbon intensity decrease. Regions corresponding to runaway emissions growth and approximate stabilization are labeled. The marginal distributions of rates of global economic growth and carbon intensity decrease are shown in blue.}
    \label{fig:growth-scatter}
\end{figure}

\begin{figure}
    \centering
    \includegraphics[width=\textwidth]{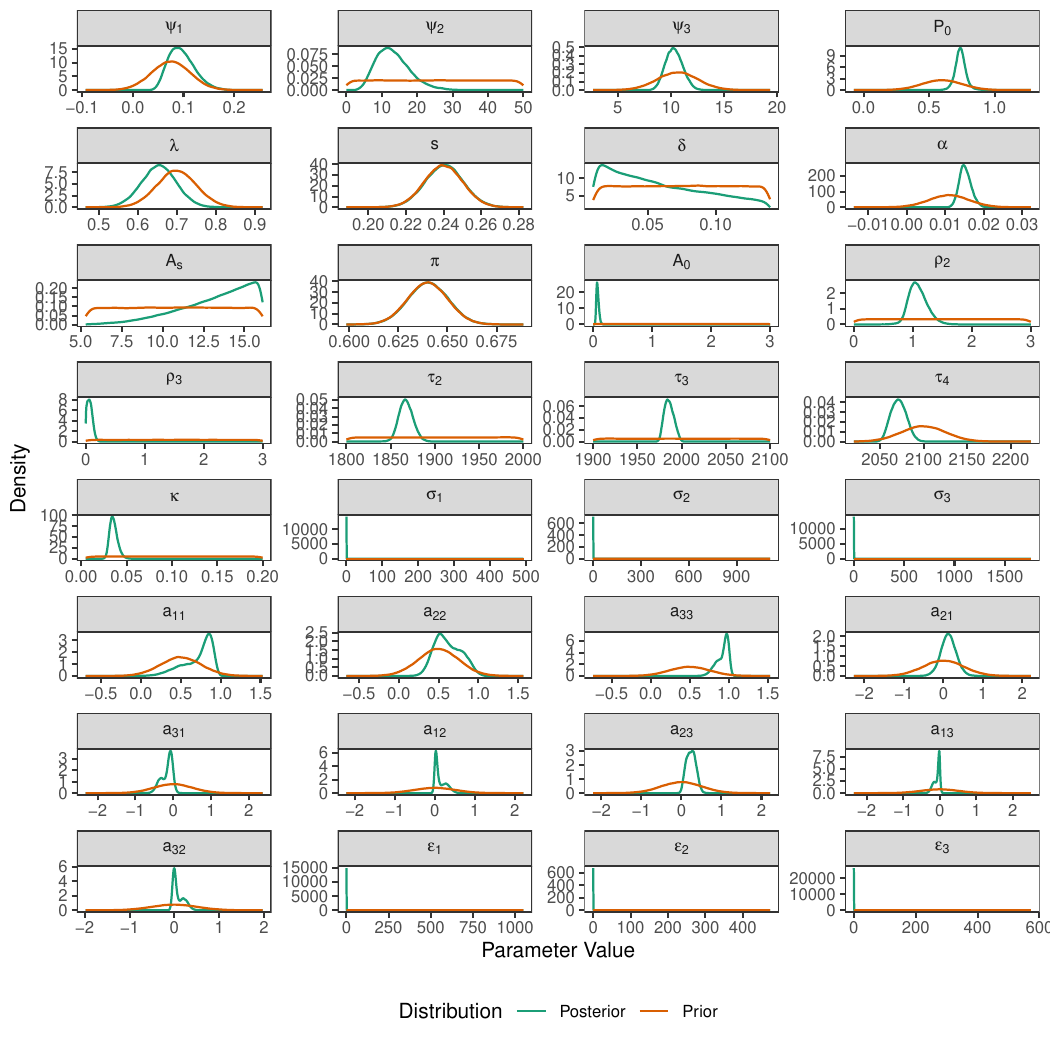}
    \caption{\textbf{Prior and posterior distributions for model parameters --} Prior and posterior distributions for the model parameters under the standard scenario. Prior distributions are in orange; posterior distributions are in green.}
    \label{fig:prior-post}
\end{figure}

\begin{figure}
    \centering
    \includegraphics[width=\textwidth]{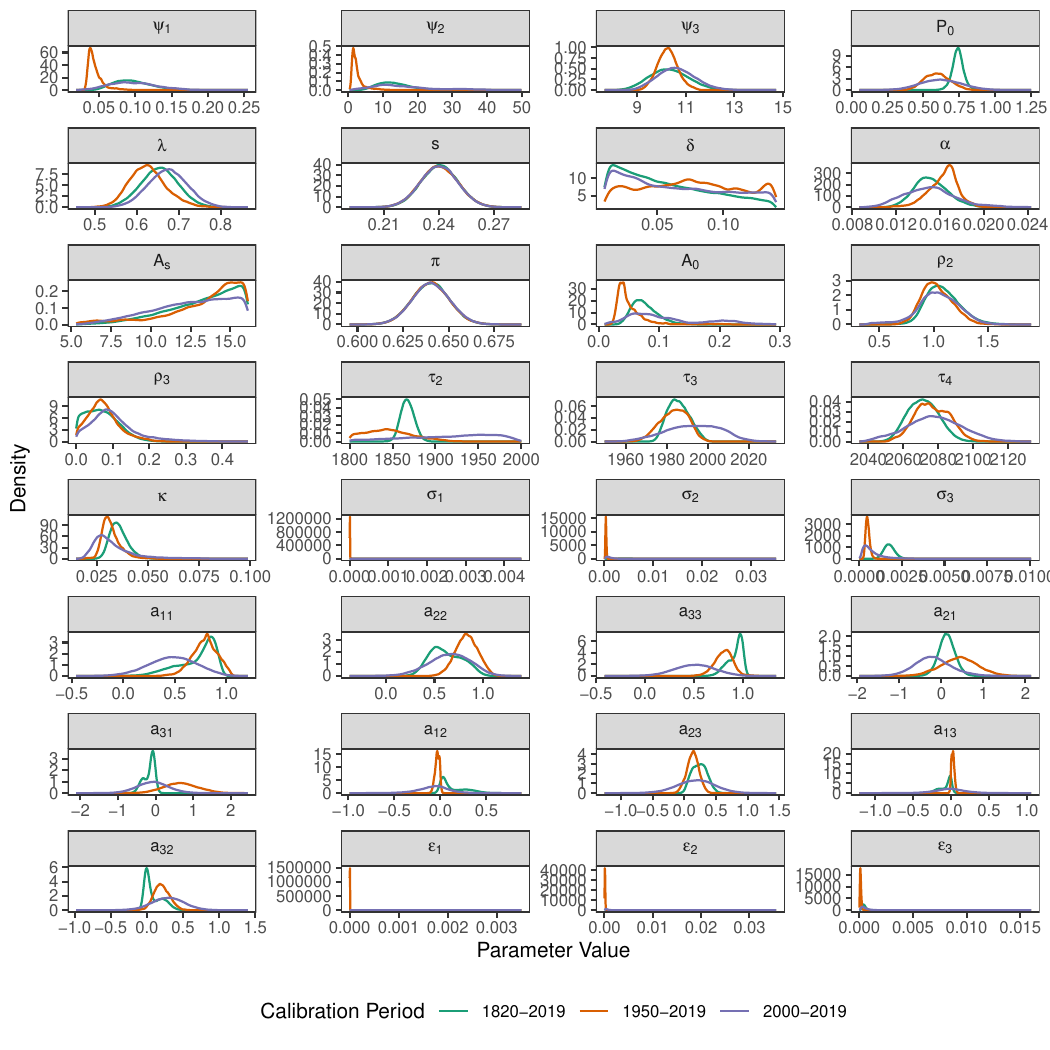}
    \caption{\textbf{Parameter posterior distributions by model scenario --} Marginal posterior distributions for the model parameters under each model scenario.}
    \label{fig:dist-calibration}
\end{figure}

\begin{figure}
    \centering
    \includegraphics[width=\textwidth,height=\textheight,keepaspectratio]{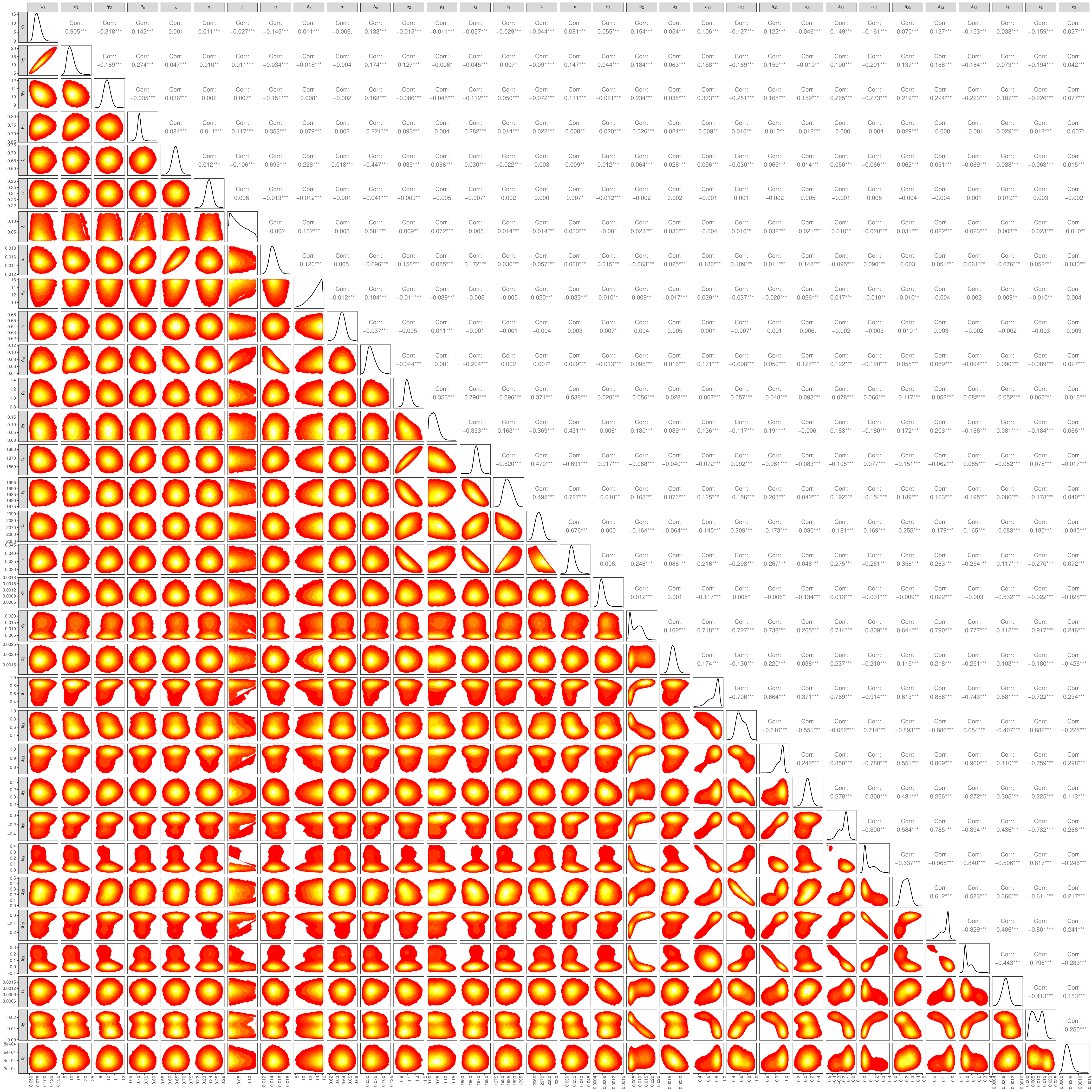}
    \caption{\textbf{Pairs plot for the standard calibration --} Pairs plot of the posterior distribution resulting from the standard calibration. Marginal distributions are plotted along the diagonal. The elements below the diagonal are heatmaps showing the joint bivariate densities. Correlation coefficients between pairs of parameters are provided above the diagonal.}
    \label{fig:pairs}
\end{figure}

\begin{figure}[hb!]
    \centering
    \includegraphics{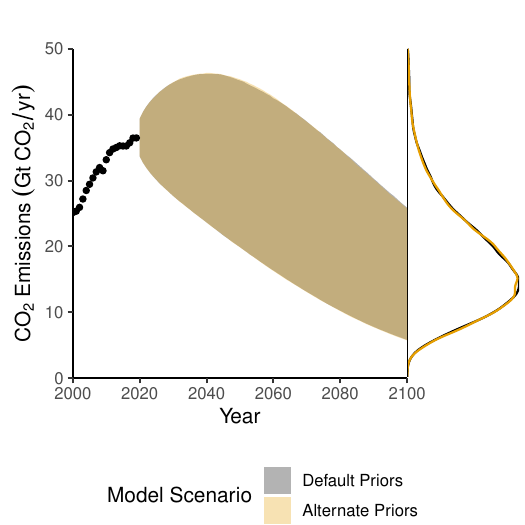}
    \caption{\textbf{Sensitivity of CO$_2$ emissions projections to prior distributions --} Projections are for the default priors (Supplemental Table 1 and Supplemental Table 2) and an alternate set of priors (Supplemental Table 3). The shaded regions are the 90\% credible intervals. Black dots are observations. The marginal distribution of projected business-as-usual CO$_2$ emissions in 2100  is shown on the right.}
    \label{fig:sens-proj}
\end{figure}